\documentclass[fleqn,usenatbib]{mnras}

\usepackage{newtxtext,newtxmath}
\usepackage{graphbox,graphicx}
\usepackage[flushleft]{threeparttable}

\usepackage[T1]{fontenc}

\DeclareRobustCommand{\VAN}[3]{#2}
\let\VANthebibliography\thebibliography
\def\thebibliography{\DeclareRobustCommand{\VAN}[3]{##3}\VANthebibliography}

\usepackage{graphicx}	
\usepackage{amsmath}	
\usepackage{siunitx}    

\title[Giant nebula around a radio-quiet quasar at $z\approx 0.6$]{The first comprehensive study of a giant nebula around a radio-quiet quasar in the $z<1$ Universe}

\author[Liu et al.]{
\newauthor
Zhuoqi (Will) Liu$^{1}${\href{https://orcid.org/0000-0002-2662-9363}{\includegraphics[scale=0.05]{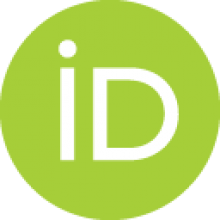}}}\thanks{E-mail: zql@umich.edu},
Sean D. Johnson$^{1}${\href{https://orcid.org/0000-0001-9487-8583}{\includegraphics[scale=0.05]{./orcid-ID.png}}},
Jennifer I-Hsiu Li$^{1, 2}${\href{https://orcid.org/0000-0002-0311-2812}{\includegraphics[scale=0.05]{./orcid-ID.png}}},
Gwen C. Rudie$^{3}${\href{https://orcid.org/0000-0002-8459-5413}{\includegraphics[scale=0.05]{./orcid-ID.png}}},
Joop Schaye$^{4}${\href{https://orcid.org/0000-0002-0668-5560}{\includegraphics[scale=0.05]{./orcid-ID.png}}},
\newauthor
Hsiao-Wen Chen$^{5}${\href{https://orcid.org/0000-0001-8813-4182}{\includegraphics[scale=0.05]{./orcid-ID.png}}},
Jarle Brinchmann$^{6}${\href{https://orcid.org/0000-0003-4359-8797}{\includegraphics[scale=0.05]{./orcid-ID.png}}},
Sebastiano Cantalupo$^{7}${\href{https://orcid.org/0000-0001-5804-1428}{\includegraphics[scale=0.05]{./orcid-ID.png}}},
Mandy C. Chen$^{5}${\href{https://orcid.org/0000-0002-8739-3163}{\includegraphics[scale=0.05]{./orcid-ID.png}}},
\newauthor
Wolfram Kollatschny$^{8}${\href{https://orcid.org/0000-0002-0417-1494}{\includegraphics[scale=0.05]{./orcid-ID.png}}},
Michael V. Maseda$^{9}${\href{https://orcid.org/0000-0003-0695-4414}{\includegraphics[scale=0.05]{./orcid-ID.png}}},
Nishant Mishra$^{1}${\href{https://orcid.org/0000-0002-9141-9792}{\includegraphics[scale=0.05]{./orcid-ID.png}}},
Sowgat Muzahid$^{10}${\href{https://orcid.org/0000-0003-3938-8762}{\includegraphics[scale=0.05]{./orcid-ID.png}}}
\\
$^{1}$Department of Astronomy, University of Michigan, 1085 S. University, Ann Arbor, MI 48109, USA\\
$^{2}$Michigan Institute for Data Science, University of Michigan, Ann Arbor, MI, 48109, USA\\
$^{3}$The Observatories of the Carnegie Institution for Science, 813 Santa Barbara Street, Pasadena, CA 91101, USA\\
$^{4}$Leiden Observatory, Leiden University, PO Box 9513, NL-2300 RA Leiden, The Netherlands\\
$^{5}$Department of Astronomy \& Astrophysics, The University of Chicago, Chicago, IL 60637, USA\\
$^{6}$Instituto de Astrofísica e Ciências do Espaço, Universidade do Porto, CAUP, Rua das Estrelas, PT4150-762 Porto, Portugal\\
$^{7}$Department of Physics, University of Milan Bicocca, Piazza della Scienza 3, I-20126 Milano, Italy\\
$^{8}$Institut für Astrophysik und Geophysik, Universität Göttingen, Friedrich-Hund Platz 1, D-37077 Göttingen, Germany\\
$^{9}$Department of Astronomy, University of Wisconsin-Madison, 475 N. Charter St., Madison, WI 53706 USA\\
$^{10}$Inter-University Centre for Astronomy and Astrophysics (IUCAA), Post Bag 4, Ganeshkhind, Pune 411 007, India
}

\date{Accepted XXX. Received YYY; in original form ZZZ}

\pubyear{2023}

\begin{document}
\label{firstpage}
\pagerange{\pageref{firstpage}--\pageref{lastpage}}
\maketitle

\begin{abstract}
We present the first comprehensive study of a giant, $\approx \! \! 70$ kpc-scale nebula around a radio-quiet quasar at $z<1$. The analysis is based on deep integral field spectroscopy with MUSE of the field of HE\,0238$-$1904, a luminous quasar at $z=0.6282$. The nebula emits strongly in $\mathrm{[O \, II]}$, $\rm H \beta$, and $\mathrm{[O \, III]}$, and the quasar resides in an unusually overdense environment for a radio-quiet system. The environment likely consists of two groups which may be merging, and in total have an estimated dynamical mass of $M_{\rm dyn}\approx 4\times 10^{13}$ to $10^{14}\ {\rm M_\odot}$. The nebula exhibits largely quiescent kinematics and irregular morphology. The nebula may arise primarily through interaction-related stripping of circumgalactic and interstellar medium (CGM/ISM) of group members, with some potential contributions from quasar outflows. The simultaneous presence of the giant nebula and a radio-quiet quasar in a rich environment suggests a correlation between such circum-quasar nebulae and environmental effects. This possibility can be tested with larger samples. The upper limits on the electron number density implied by the [O\,II] doublet ratio range from $\log(n_{\rm e, [O \, II]} / \mathrm{cm^{-3}}) < 1.2$ to $2.8$. However, assuming a constant quasar luminosity and negligible projection effects, the densities implied from the measured line ratios between different ions (e.g., [O\,II], [O\,III], and [Ne\,V]) and photoionization simulations are often 10$-$400 times larger. This large discrepancy can be explained by quasar variability on a timescale of $\approx 10^4{-}10^5$ years.
\end{abstract}




\section{Introduction}
Galaxy evolution is a complex process that involves gas inflows and outflows thought to control star formation and black hole growth \citep[for a review, see][]{2017ARA&A..55...59N}. Observations of interstellar medium (ISM) gas masses and star formation rates suggest that massive star-forming galaxies have an ISM depletion timescale much smaller than the age of the Universe at $z < 3$ \citep{2012ARA&A..50..531K, 2013ApJ...768...74T}. This can be explained if galaxies accrete gas from external sources to maintain their star-forming activity and black hole growth \citep[though see][]{2011ApJ...734...48L}. At the same time, the ISM of galaxies can lose gas through various processes including stellar \citep[for a review, see][]{2018Galax...6..114Z} and AGN feedback \citep[for a review, see][]{2012ARA&A..50..455F}, ram pressure stripping \citep[e.g.,][]{2006ApJ...647..910H}, and tidal interactions with neighboring galaxies \citep[e.g.,][]{2016MNRAS.461.2630M}. Therefore, observations of the physical conditions, kinematics, and distribution of gas around galaxies can provide insights into the mechanisms governing galaxy formation and evolution. For these reasons, observations of the gaseous cosmic ecosystems of galaxies were highlighted as a key long-term priority by the \textcolor{black}{2020 Decadal Survey for Astronomy and Astrophysics} \citep{2021pdaa.book.....N}.

\textcolor{black}{The properties of} gas flows around galaxies, including their morphology and kinematics, can be directly traced by observations of giant gas nebulae with state-of-the-art wide-field integral field spectrographs (IFSs) such as \textcolor{black}{the Multi-Unit Spectroscopic Explorer (MUSE; \citealt{2010SPIE.7735E..08B}) and the Keck Cosmic Web Imager (KCWI; \citealt{2010SPIE.7735E..0MM})}. At $z > 2$, systematic IFS surveys around radio-quiet quasars discovered ubiquitous giant H\,I Ly$\alpha$ nebulae \citep[e.g.,][]{2014Natur.506...63C, 2016ApJ...831...39B, 2019ApJS..245...23C, 2020ApJ...894....3O, 2021MNRAS.503.3044F, 2021MNRAS.502..494M}. More recently, a study of the ionization states of one of these nebulae found that the gas has a surprisingly large density for halo-scale emission or a very broad density distribution \citep{2019MNRAS.483.5188C}. However, due to redshifting of optical emission lines into the infrared, surface brightness dimming, and the faintness of galaxies at high redshift, more fully characterizing these $z > 2$ nebulae is time-consuming even with large space- or ground-based telescopes \citep[though see][]{2023MNRAS.519.5099L}.

At low redshift, on the other hand, non-resonant emission lines such as $\rm [O \, II]$, $\rm H \beta$, and $\rm [O \, III]$ are available at optical wavelengths, and collecting galaxy spectra is less expensive. The power of IFSs enabled the discoveries of giant nebulae around starburst galaxies, galaxy groups, and quasars \citep[e.g.,][]{2018A&A...609A..40E, 2019A&A...631A.114B, 2019ApJ...878L..33C, 2019Natur.574..643R, 2021MNRAS.507.4294Z, 2021ApJ...909..151B, 2022A&A...663A..11L, 2023arXiv230209087D}, arising from outflows, interactions, and filamentary accretion. These low redshift nebulae provide an opportunity to study the physical conditions and the processes that may produce giant nebulae at higher redshift. Most published studies of giant nebulae around $z < 1$ quasars have focused on radio-loud systems \citep{2018ApJ...869L...1J, 2021MNRAS.505.5497H, 2022ApJ...940L..40J}, which represent a small fraction of the general quasar population \citep[e.g.,][]{1989AJ.....98.1195K}. Furthermore, clustering measurements indicate that radio-loud quasars typically reside in massive galaxy groups with halo masses of $M \sim 10^{13}\  {\rm M_{\odot}}$ while the halo masses of more common radio-quiet systems are approximately five times lower on average \citep[e.g.,][]{2009ApJ...697.1656S}. This mass miss-match and the possibility \textcolor{black}{of} radio jet feedback make the comparison between low-redshift giant nebulae around radio-loud quasars and high-redshift radio-quiet ones difficult.

Recently, \citet{2023MNRAS.518.2354C} demonstrated the existence of giant nebulae around two radio-quiet quasars as part of a study focused on turbulence using the observed velocity structure function. In this paper, we present the first \textcolor{black}{comprehensive characterization of a giant nebula and associated galaxy environment around a radio-quiet quasar at $z < 1$, HE\,0238$-$1904}. \textcolor{black}{Recently, this nebula was independently discovered and reported by \cite{2023ApJ...943L..25Z}. However, our interpretation of the system differs substantially from the one presented by \cite{2023ApJ...943L..25Z} due to adoption of a significantly different quasar systemic redshift}. \textcolor{black}{In particular, \cite{2023ApJ...943L..25Z} adopted a Mg\,II emission-based redshift of $z=0.631$ from the Hamburg/ESO Survey of bright Quasars \citep[][]{2000A&A...358...77W}. On the other hand, we adopt a redshift estimate of $z=0.6282$ based on the [O\,II] emission-line centroid measured in the spectrum of the quasar extracted from the same MUSE dataset used to measure the kinematics of the giant nebula.} The paper is organized as follows: In Section \ref{OD}, we discuss the observations, data reduction, and processing. In Section \ref{ME}, we describe our measurements and investigate the group environment and giant nebula properties. In Section \ref{Dis}, we investigate the origin of the nebula and the physical conditions of the gas. In Section \ref{SC}, we summarize our findings and discuss their implications.

Throughout the paper, we adopt a flat $\Lambda$ cosmology with $\Omega_{\rm m}=0.3$, $\Omega_{\rm \Lambda}=0.7$, and $H_{0} = 70 \, \rm km \, s^{-1} Mpc^{-1}$. All magnitudes are given in the AB system unless otherwise stated.

\section{Observations and Data}
\label{OD}
\textcolor{black}{The $z\approx0.63$ quasar HE\,0238$-$1904 has high-quality archival UV \textcolor{black}{\it HST} absorption spectra used to study the CGM of the Milky Way \citep{2019ApJ...871...35Z, 2021ApJ...912....8B} and distant galaxies \citep{2018MNRAS.476.4965M, 2018ApJ...866...33L} in addition to a highly ionized, fast outflow from the quasar itself \citep{2012MNRAS.424L..59M, 2013MNRAS.436.3286A}.} To identify faint foreground galaxies in the quasar field, we observed it with MUSE as part of the Quasar-field Blind Emitter Survey (MUSE-QuBES; \citealt{2020MNRAS.496.1013M, 2023arXiv230316933D}) on the Very Large Telescope (VLT; PI: J. Schaye, PID: 094.A-0131(B) \& 096.A-0222(A)).
MUSE is an integral-field spectrograph on the UT4 VLT with a field of view (FoV) of $1'\times 1'$ and a spaxel size of $0.2''$ in wide-field mode (WFM). MUSE covers the spectral range between $4750\, \text{\AA}$ to $9350 \, \text{\AA}$ and a resolution of $R\sim3000$. The MUSE observations are centered near the quasar sightline, and we obtained eleven exposures collected between November 18th, 2014 and February 2nd, 2016 with a total exposure time of 8.75 hr with median seeing full-width-at-half-maximum (FWHM) conditions of $0.7''$. At the redshift of HE\,0238$-$1904, the MUSE FoV corresponds to a projected size of $\approx 400$ proper kpc (pkpc) on a side, and the spectral coverage includes emission lines such as [O\,II], H$\beta$, and [O\,III]. \textcolor{black}{These emission lines} enable sensitive studies of any ionized nebulae and galaxies in the quasar's environment.

\begin{figure*}
    \centering
    \includegraphics[width=\linewidth]{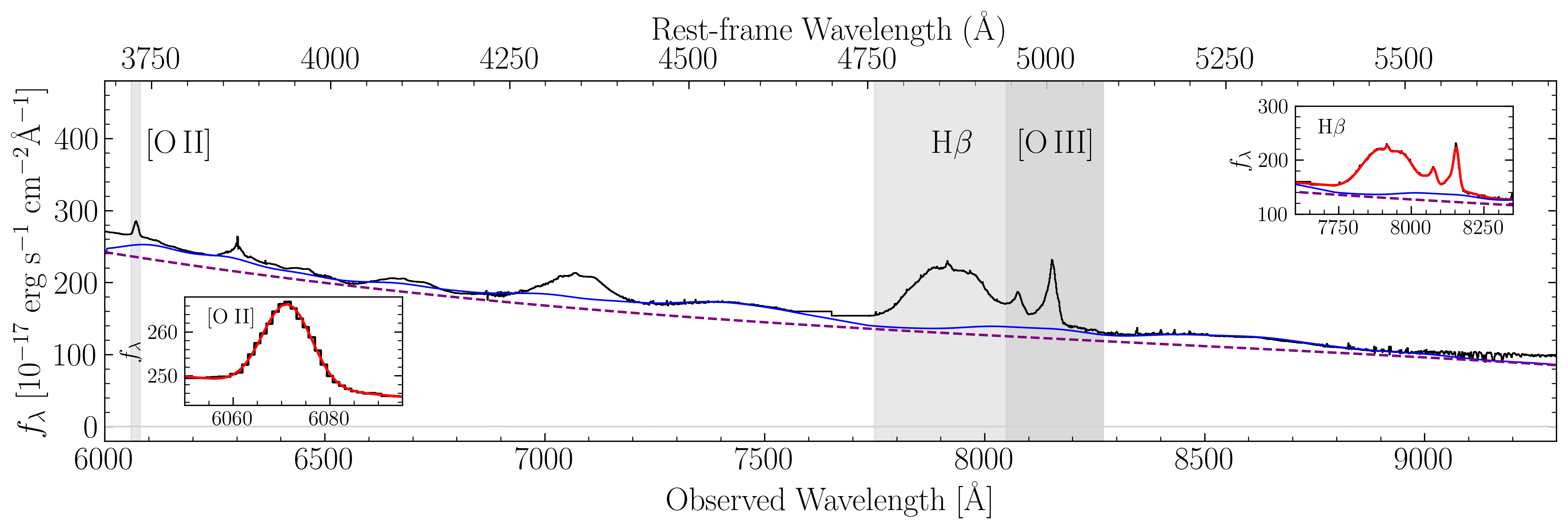}
    \caption{MUSE spectrum of HE\,0238$-$1904 overplotted with best-fit models. The MUSE spectrum is shown as a solid black line, the power-law continuum model is shown as a dashed purple line, and the iron template model is shown using a solid blue line. The bottom left inset panel shows the $\mathrm{[O \, II]}$ line emission with the best-fit continuum+line model shown in red. The top right inset panel shows the $\rm H\beta$ and [O\,III] emission with the best-fit shown in red. We measured the systemic redshift of the quasar from the [O\,II] doublet, and inferred the black hole mass from the $\rm H\beta$ broad component and the continuum luminosity at 5100\AA\ as described in detail in Section \ref{MEQP}.}
    \label{fig:qso}
\end{figure*}

To ensure robustness of results, we analyzed the MUSE data reduced through three independent pipelines including CubEx \citep{2019MNRAS.483.5188C}, the MUSE GTO team pipeline \citep{2014ASPC..485..451W}, and the ESO reduction pipeline \citep{2012SPIE.8451E..0BW} and found consistent results with all three. All three pipelines include bias subtraction, flat fielding, wavelength calibration, geometric calibration, sky subtraction, flux calibration, and stacking of exposures. For the ESO reductions, we obtained the final, stacked datacube from the ESO Science Archive and performed additional post-processed sky subtraction with the Zurich Atmosphere Purge package (ZAP; \citealt{2016MNRAS.458.3210S}). For simplicity, we converted the air wavelengths delivered by the three pipelines to vacuum.

To enable more sensitive and higher angular resolution photometric measurements of galaxies in the quasar field, we also obtained an image from the Advanced Camera for Surveys (ACS) on the \textit{Hubble Space Telescope} (\textit{HST}) with the F814W filter (PI: L. Straka, PID: 14660) with a total exposure time of $2182$ seconds split between four dithered exposures. We obtained the reduced, stacked image from the Barbara A. Mikulski Archive for Space Telescopes (MAST). In addition, to measure the UV luminosity of the quasar, we obtained the archival UV spectrum from the Cosmic Origins Spectrograph (COS; \citealt{2012ApJ...744...60G}) from MAST. \textcolor{black}{The spectrum} consists of a total exposure time of 14400 seconds and 7496 seconds in the G130M and G160M gratings, respectively (PI: J. Green and S. Penton, PID: 11541 and 12505). We reduced and coadded the COS spectrum following procedures outlined in \citet{2015MNRAS.449.3263J, 2020MNRAS.497..498C}.

\subsection{Quasar Light Subtraction}
HE\,0238$-$1904 has a Gaia \citep{2018A&A...616A...1G} $G$-band magnitude of $m_{G}=15.2$, and this brightness combined with the broad wings of the MUSE point spread function (PSF) causes contamination of nearby \textcolor{black}{galaxy spectra} with quasar light. This contamination includes both continuum and line emission due to the unresolved narrow-line region in the nucleus. To study faint extended emission, we removed the contamination by performing quasar light subtraction as described in \citet{2021MNRAS.505.5497H}. In summary, our method of quasar light subtraction does not rely on PSF measurements. Instead, it uses spectral information and the fact that quasars and galaxies have different spectral energy distributions \citep[see also][]{2017ApJ...850...40R, 2023MNRAS.518.2354C}.

In ground-based observations, the Earth's atmosphere scatters bluer photons more than redder ones so that the PSF is wider at bluer wavelengths. The differential scattering makes the spectral slope observed in a spaxel depend on the angular separation from the quasar with steeper (shallower) slopes further from (closer to) the quasar centroid. To account for this, we used a two\textcolor{black}{-}component non-negative matrix factorization (NMF; \citealt{2007AJ....133..734B, 2018ApJ...852..104R}) of the quasar light, with one component having a shallow slope and a second having a steep slope. \textcolor{black}{Adding additional a third or fourth NMF component(s) did not noticeably improve the results.} In general, the spectrum for each spaxel near the quasar has some light from the quasar and potentially nearby galaxies as well. To subtract quasar light while avoiding subtraction of galaxy light, we fit each spaxel with a linear combination of the two quasar non-negative components and the first two \textcolor{black}{Sloan Digital Sky Survey-Baryon Oscillation Spectroscopic Survey (SDSS-BOSS)} galaxy eigenspectra \citep{2012AJ....144..144B} and then subtracted the quasar component of the model. \textcolor{black}{Unlike with some other systems \citep[e.g.,][]{2018ApJ...869L...1J}, the host of HE\,0238$-$1904 does not exhibit bright, extended starlight, so the contribution inferred by the galaxy model was not significant.}

\section{Measurements and Environment}
\label{ME}
\subsection{Quasar Properties}
\label{MEQP}
HE\,0238$-$1904 is a luminous, radio-quiet quasar \citep{2006A&A...455..773V, 2013MNRAS.436.3286A}. To ensure self-consistent measurements of the quasar properties, we estimated its redshift, luminosity, and black hole mass using the MUSE spectrum extracted via \texttt{MPDAF} \citep{2016ascl.soft11003B} with a $r=3''$ aperture. To measure the systemic redshift of the quasar, we fit the $\rm [O \, II] \lambda \lambda 3727, 3729$ doublet with a Gaussian profile following \citet{2010MNRAS.405.2302H} and found $z = 0.6282 \pm 0.0002$, where the uncertainty represents the scatter between the $\rm [O \, II]$ centroid and stellar absorption lines of SDSS quasars at similar redshift. This redshift is $\approx +500\ {\rm km\,s^{-1}}$ \ from a previously reported Mg\,II based estimate from \cite{2000A&A...358...77W}. Even so, a more recent Mg\,II based redshift of $z=0.628$ from \cite{2016AJ....152...25M} confirms our [O\,II]-based redshift estimate. In general, quasar redshifts measured from the [O\,II] doublet are more accurate than those measured from broad-lines like Mg\,II, as we argue in Section \ref{subsection:origin}.

In addition, we estimated the bolometric luminosity and the black hole mass of HE\,0238$-$1904 by fitting the extracted MUSE spectrum with the Python QSO fitting code (\texttt{PyQSOFit}; \citealt{2019MNRAS.482.3288G}). \texttt{PyQSOFit} fits a quasar's spectrum with a combination of a power-law continuum, $\mathrm{Fe \, II}$ template, and sets of Gaussian line profiles for both the broad- and narrow-lines. We modelled the $\rm H\beta$ and [O\,III] spectral region with the continuum components, three Gaussian profiles for the broad H$\beta$, and two for the narrow H$\beta$ and [O\,III]. From the fit, we computed a monochromatic luminosity at $5100$\AA \, of $\lambda L_{5100} \approx 1.6 \times 10^{46} \rm\ erg \, s^{-1}$ and a bolometric luminosity of $L_{\rm bol} \approx 1.7 \times 10^{47} \, \rm erg \, s^{-1}$ using the bolometric correction factor from \cite{2006ApJS..166..470R}. Finally, we inferred a black hole mass of $M_{\rm BH}\approx 10^{9.8}\ {\rm M_{\odot}}$ using the single-epoch virial theorem-based approach from  \cite{2006ApJ...641..689V}. \textcolor{black}{Following \citet{2013ARA&A..51..511K}, this black hole mass corresponds to a stellar mass of $M_{*}\approx 10^{12.0}\ {\rm M_{\odot}}$ for the host galaxy, but we caution this stellar mass may be significantly overestimated due to uncertainty in single-epoch virial theorem-based black hole masses and observed scatter in the black hole mass-stellar mass relation. For example, if the true black hole mass is $1\sigma$ below the mean single-epoch virial theorem estimate, and the stellar mass is $1\sigma$ below the estimate from the black hole mass-stellar mass relation, the inferred stellar mass would be $M_{*}\approx 10^{11.4}\ {\rm M_{\odot}}$. Furthermore, the single-epoch virial theorem-based relation used here is not calibrated for quasars as luminous as HE\,0238$-$1904, which may drive disk wind, erroneously inflating the black hole mass estimate.} The fitted quasar spectrum is shown in Figure \ref{fig:qso}.

\begin{figure*}
    \centering
    \includegraphics[scale=1.0]{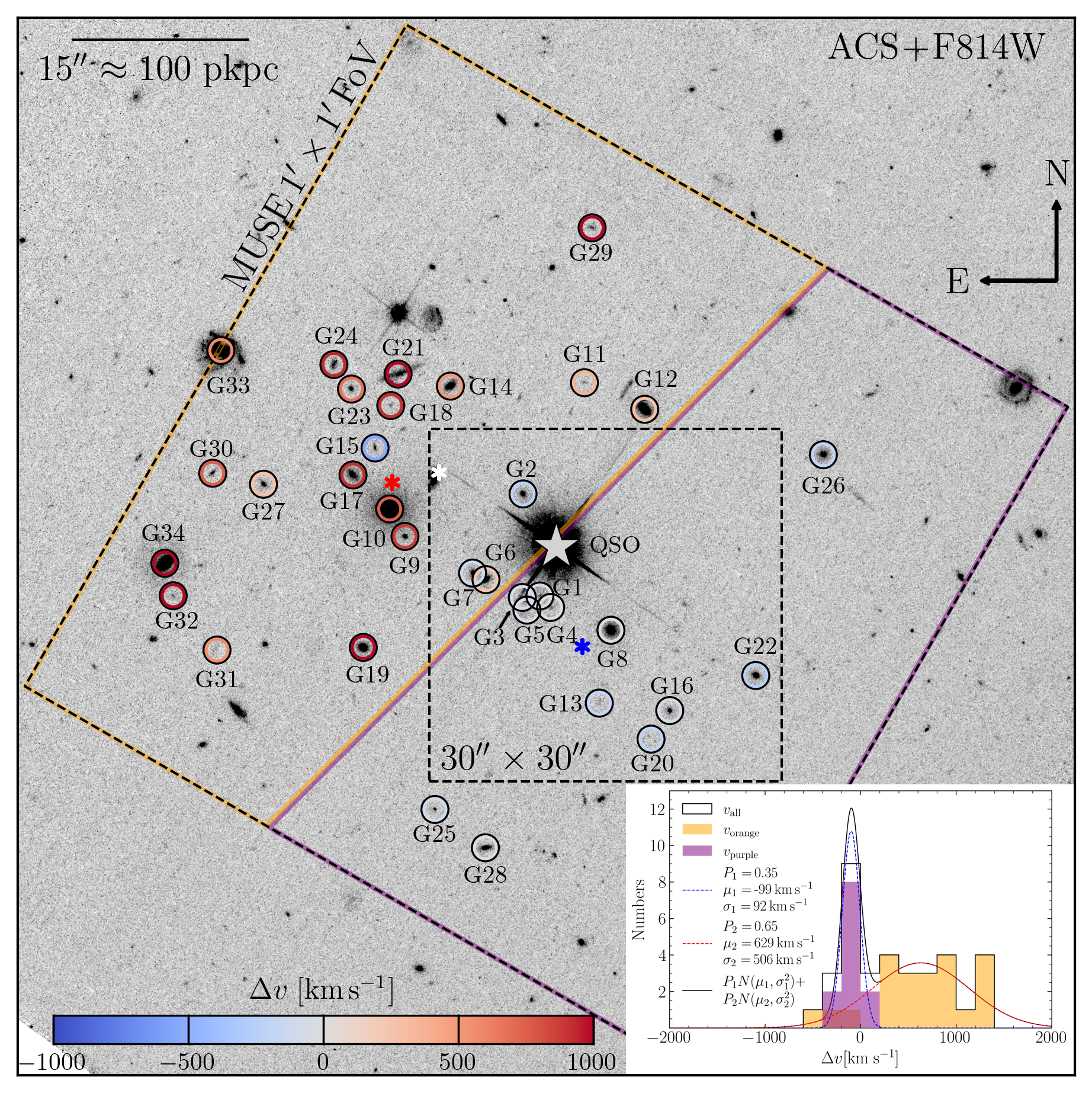}
    \caption{\textit{HST} ACS+F814W image of the field of HE 0238-1904. The full image has a FoV of $1.5' \times 1.5'$. The larger dashed box shows the $1' \times 1'$ MUSE FoV. The smaller dashed box marks the $30'' \times 30''$ region displayed in Figure \ref{fig:gas}. The LOS velocities of galaxies relative to the quasar are denoted with outlining colors and the corresponding colorbar is shown on the bottom left. The histogram in the bottom right inset panel shows the velocity distribution of galaxies where galaxies in both orange and purple outlined regions are plotted separately. \textcolor{black}{We note that the orange and purple regions and corresponding histograms are only for visualization. The two-Gaussian fitting of the velocity distribution does not rely on any spatial information.} Galaxies in the quasar host environment are labeled with black circles and labeled by their IDs. The approximate stellar mass weighted group center is marked with a white asterisk while the weighted centers of the richer, redshifted group and less rich, blueshifted group are marked with red and blue asterisks, respectively. Based on spatial distribution and kinematics, HE\,0238$-$1904 resides in a massive, rich environment potentially consisting of two galaxy groups which may be merging.}
    \label{fig:dis}
\end{figure*}

\subsection{Galaxy Measurements and Properties}
\label{GMP}
To study the environment of HE\,0238$-$1904, we \textcolor{black}{conducted} a galaxy survey by first identifying all continuum sources in MUSE and the ACS$+$F814W image. \textcolor{black}{We identified continuum sources by running \texttt{Source Extractor} (\texttt{SE}; \citealt{1996A&AS..117..393B}) on a median MUSE white light image and the {\it HST} image separately. To ensure completeness, we also added sources based on visual inspection.} Typically, sources are missing from MUSE due to biased background estimation caused by bright objects in the field or due to blending. Based on the background sky standard deviation and source counts in the ACS$+$F814W image, the imaging catalog is complete for objects brighter than $m_{\rm F814W}\approx26\!-\!27 $, depending on angular size.

\begin{figure*}
    \centering
    \includegraphics[scale=0.7]{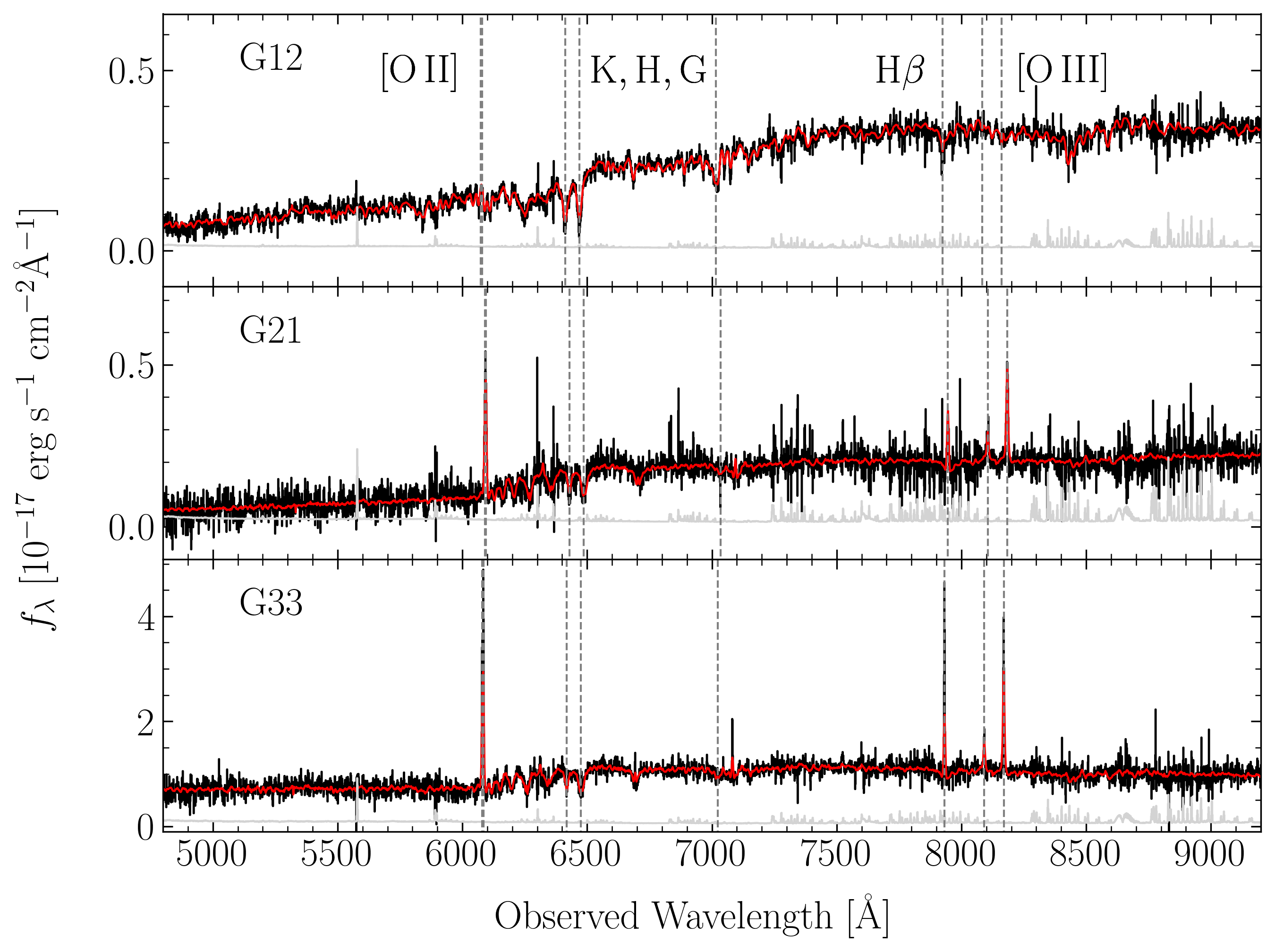}
    \caption{MUSE galaxy spectra with the best-fit spectral models. The MUSE spectrum is shown by a solid black line. The uncertainty is shown by a solid grey line. The best-fit model used for redshift measurement is shown by a solid red line.}
    \label{fig:gal}
\end{figure*}

For each identified object, we extracted a MUSE spectrum with \texttt{MPDAF} with a circular aperture of $r=0.7''$, which is roughly the size of the MUSE seeing FWHM. \textcolor{black}{The choice of this modest aperture may result in some wavelength dependent aperture losses but helps increase S/N for redshift estimation.}  We then fit each spectrum as a linear combination of SDSS galaxy eigenspectra as described in \citet{2021MNRAS.505.5497H} to measure the source redshift. In summary, we computed the best-fit linear combination on a grid from $z=0$ to $z=1$ with a step size of $\Delta z = 0.0001$ and recorded the goodness-of-fit statistic ($\chi^2$) over the entire grid. We adopted the redshift with the minimum global $\chi^2$ as our initial solution. We then visually inspected each best-fit model to ensure robustness and assigned the redshift quality. For galaxies with both emission and absorption lines, we masked out strong emission lines and measured the redshift based on stellar absorption features when possible to avoid a potential bias in redshift from large-scale nebulae in the field (which may not be closely associated with the galaxies in question). Finally, we classified our confidence in the redshift measurements based on the number of the detected spectral features. \textcolor{black}{All of the galaxies in the quasar environment have two or more spectral features except for G11 and G18}. According to \citet{2021MNRAS.505.5497H}, the uncertainty in galaxy redshifts measured in MUSE spectra with these techniques is $\sigma \approx \rm 20 \, km \, s^{-1}$. Comparing the continuum source catalog and the corresponding redshift measurements, the redshift survey is approximately $100\%$ complete for sources brighter than $m_{\rm F814W} \! \approx \! 24$ and approximately $95\%$ complete for those brighter than $m_{\rm F814W} \! \approx \! 25$. \textcolor{black}{For comparison, an $L_*$ galaxy at $z \approx 0.6$ has $m_{\rm F814W} \! \approx \! 20.6$ assuming the luminosity function from \cite{2007ApJ...665..265F}.} The high completeness of the galaxy survey at faint magnitudes enables us to study the origins of nebulae, even if they arise from interactions involving relatively faint dwarf galaxies.

\begin{table*}
	\centering
	\caption{Summary of Galaxies in the Field of HE\,0238$-$1904 at \textit{$z \approx z_{\rm{QSO}}$}.}
	\label{tab:summary_galaxies}
	\begin{threeparttable}
	\begin{tabular}{cccccccccrrrr} 
		\hline
		ID & R.A.\tnote{a} & Decl.\tnote{b} & \textit{z}\tnote{c} & \textit{$m_{\mathrm{F814W}}$}\tnote{d} & \textit{$M_{B}$}\tnote{e} &
		\textit{K}-correction & \multicolumn{1}{c}{D4000} & $A_V$ & \multicolumn{1}{c}{$\log (M_*/{\rm M_\odot})$\tnote{f}} & \multicolumn{1}{c}{$\Delta \theta$\tnote{g}} & \multicolumn{1}{c}{\textit{d}\tnote{h}} & \multicolumn{1}{c}{$\Delta$\textit{v}\tnote{i}} \\
		& (J2000) & (J2000) & & (AB) & (AB) & template & & (mag) & & ($''$) & (pkpc) & ($\mathrm{km\ s^{-1}}$) \\
		\hline
		Host & 02:40:32.58 & \textcolor{black}{$-$18:51:51.4} & 0.6282 & ... & ... & ... & ... & ... &  ... & 0.0 &   0.0 & 0\\
    	G1  & 02:40:32.63 & \textcolor{black}{$-$18:51:55.8} & 0.6278 & 24.3 & $-$17.5 & S0 & $1.26 \pm 0.57$ & ... & \textcolor{black}{9.3}\tnote{j} &  4.4 &  30.4 & -76\\
        G2  & 02:40:32.73 & \textcolor{black}{$-$18:51:47.1} & 0.6270 & 23.3 & $-$18.5 & S0 & $1.56 \pm 0.08 $ & 0.1 & \textcolor{black}{9.5} &  4.8 &  32.7 & -224\\
        G3\tnote{k} & 02:40:32.74 & \textcolor{black}{$-$18:51:55.9} & 0.6280 & 23.8 & $-$18.3 & Irr & ... & ... & \textcolor{black}{9.6}\tnote{j} & 5.0 & 34.3 & -40\\
		G4  & 02:40:32.57 & \textcolor{black}{$-$18:51:56.7} & 0.6284 & 24.9 & $-$17.3 & Irr & $1.05 \pm 0.07$ & 0.2 & \textcolor{black}{8.3} & 5.4 & 36.7 & +34\\
		G5  & 02:40:32.71 & \textcolor{black}{$-$18:51:57.0} & 0.6280 & 25.2 & $-$17.0 & Irr & $0.64 \pm 0.08$ & 0.1 & \textcolor{black}{7.4} &  5.9 &  40.1 & -40\\
		G6  & 02:40:32.96& \textcolor{black}{$-$18:51:54.4} & 0.6295 & 22.4 & $-$19.4 & S0 & $1.35 \pm 0.02$ & 0.1 & \textcolor{black}{10.1} & 6.1 & 41.5 & +237\\
		G7  & 02:40:33.04 & \textcolor{black}{$-$18:51:53.8} & 0.6275 & 23.8 & $-$18.0 & S0 & $1.30 \pm 0.04$ & 0.0 & \textcolor{black}{9.3} & 6.9 & 46.9 & -132\\
		G8  & 02:40:32.21 & \textcolor{black}{$-$18:51:58.7} & 0.6284 & 21.8 & $-$20.0 & S0 & $1.62 \pm 0.02$ & 0.2 & \textcolor{black}{10.4} & 9.1 & 61.9 & +34\\
        G9  & 02:40:33.44 & \textcolor{black}{$-$18:51:50.7} & 0.6330 & 23.8 & $-$18.1 & S0 & $1.49 \pm 0.05$ & 0.2 & \textcolor{black}{9.7} &  12.2 &  82.2 & +882\\
        G10  & 02:40:33.53 & \textcolor{black}{$-$18:51:48.4} & 0.6323 & 20.0 & $-$21.9 & S0 & $1.71 \pm 0.01$ & 0.8 & \textcolor{black}{11.5} & 13.8 & 94.3 & +753\\
        G11  & 02:40:32.37 & \textcolor{black}{$-$18:51:37.6} &0.6302 & ... & ... & ... & ... & ... &\textcolor{black}{...} & 14.1 & 96.3 & +360\\ 
		G12  & 02:40:32.00 & \textcolor{black}{$-$18:51:39.9} & 0.6297 & 21.4 & $-$20.4 & S0 & $1.64 \pm 0.02$ & 0.2 & \textcolor{black}{10.6} & 14.1 & 96.5 & +274\\
		G13  & 02:40:32.28 & \textcolor{black}{$-$18:52:04.9} & 0.6272 & ... & ... & ... & ... & ... & \textcolor{black}{...} & 14.2 & 97.0 & -187\\
        G14  & 02:40:33.17 & \textcolor{black}{$-$18:51:37.9} & 0.6310 & 22.6 & $-$19.2 & S0 & $1.37 \pm 0.03$ & 0.7 & \textcolor{black}{10.0} & 15.8 & 108.0 & +513\\
        G15  & 02:40:33.62 & \textcolor{black}{$-$18:51:43.2} & 0.6253 & 24.8 & $-$17.0 & S0 & $1.99 \pm 0.22$ & 0.4 & \textcolor{black}{9.0} & 16.8 & 115.0 & -537\\
        G16  & 02:40:31.85 & \textcolor{black}{$-$18:52:05.5} & 0.6279 & 23.8 & $-$18.0 & S0 & $1.98 \pm 0.16$ & 1.1 & \textcolor{black}{9.5} & 17.5 & 119.8 & -58\\
        G17  & 02:40:33.75 & \textcolor{black}{$-$18:51:45.5} & 0.6332 & 22.7 & $-$19.1 & S0 & $1.57 \pm 0.03$ & 0.6 & \textcolor{black}{10.1} & 17.6 & 120.3 & +919\\
		G18  & 02:40:33.53 & \textcolor{black}{$-$18:51:39.6} & 0.6332 & ... & ... & ... & ... & ... & \textcolor{black}{...} & 17.9 & 121.9 & +922\\
		G19  & 02:40:33.69 & \textcolor{black}{$-$18:52:00.1} & 0.6358 & 22.2 & $-$19.7 & S0 & $1.60 \pm 0.02$ & 0.4 & \textcolor{black}{10.3} & 18.0 & 122.9 & +1398\\
		G20  & 02:40:31.97 & \textcolor{black}{$-$18:52:07.9} & 0.6271 & ... & ... & ... & ... & ... & \textcolor{black}{...} & 18.8 & 128.1 & -205\\
        G21  & 02:40:33.48 & \textcolor{black}{$-$18:51:36.9} & 0.6341 & 22.1 & $-$19.7 & S0 & $1.26 \pm 0.02$ & 1.4 & \textcolor{black}{10.3} & 19.3 &  131.8 & +1084\\
        G22  & 02:40:31.34 & \textcolor{black}{$-$18:52:02.5} & 0.6268 & 23.0 & $-$18.9 & S0 & $1.66 \pm 0.05$ & 0.5 &\textcolor{black}{10.1} & 20.9 & 142.8 & -261\\
        G23  & 02:40:33.76 & \textcolor{black}{$-$18:51:38.2} & 0.6319 & 24.4 & $-$17.6 & S0 & $1.62 \pm 0.11$ & 0.6 & \textcolor{black}{9.5} & 21.3 & 145.5 & +679\\
		G24  & 02:40:33.87 & \textcolor{black}{$-$18:51:36.1} & 0.6333 & 23.6 & $-$18.5 & Scd & $1.07 \pm 0.04$ & 1.8 & \textcolor{black}{9.8} & 23.8 & 162.4 & +937\\
        G25  & 02:40:33.26 & \textcolor{black}{$-$18:52:13.9} & 0.6277 & 25.5 & $-$16.3 & S0 & $1.46 \pm 0.17$ & 1.8 & \textcolor{black}{8.8} & 24.5 & 167.5 & -95\\
        G26  & 02:40:30.93 & \textcolor{black}{$-$18:51:43.7} & 0.6272 & 23.0 & $-$18.8 & S0 & $1.66 \pm 0.05$ & 0.6 & \textcolor{black}{9.9} & 24.7 & 168.3 & -187\\
        G27  & 02:40:34.29 & \textcolor{black}{$-$18:51:46.3} & 0.6297 & 23.7 & $-$18.1 & S0 & $1.30 \pm 0.06$ & 0.5 & \textcolor{black}{9.5} & 24.8 & 169.0 & +274\\
		G28  & 02:40:32.96 & \textcolor{black}{$-$18:52:17.2} & 0.6282 & 23.0 & $-$19.1 & Scd & $1.07 \pm 0.02$ & 1.0 & \textcolor{black}{9.5} & 26.4 & 180.0 & -3\\
        G29  & 02:40:32.32 & \textcolor{black}{$-$18:51:24.4} & 0.6357 & 24.5 & $-$17.8 & Irr & $0.83 \pm 0.06$ & 1.4 & \textcolor{black}{8.3} &  27.2 &  185.6 & +1379\\
        G30  & 02:40:34.59 & \textcolor{black}{$-$18:51:45.3} & 0.6323 & 24.5 & $-$17.6 & Scd & $0.96 \pm 0.06$ & 0.1 & \textcolor{black}{8.9} & 29.2 & 199.2 & +753\\
        G31  & 02:40:34.57 & \textcolor{black}{$-$18:52:00.4} & 0.6312 & ... & ... & ... & ... & ... & \textcolor{black}{...} & 29.6 & 201.9 & +550\\
        G32  & 02:40:34.83 & \textcolor{black}{$-$18:51:55.7} & 0.6354 & 24.8 & $-$17.1 & S0 & $1.25 \pm 0.08$ & 0.0 & \textcolor{black}{9.0} & 32.2 &  219.9 & +1324\\
        G33  & 02:40:34.55 & \textcolor{black}{$-$18:51:34.9} & 0.6313 & 20.0 & $-$22.1 & Scd & $1.17 \pm 0.01$ & 0.4 & \textcolor{black}{10.8} &  32.4 &  220.9 & +569\\
        G34  & 02:40:34.88 & \textcolor{black}{$-$18:51:53.0} & 0.6349 & 20.6 & $-$21.2 & S0 & $1.55 \pm 0.01$ & 0.4 & \textcolor{black}{11.2} &  32.7 &  222.9 & +1232\\
		\hline \\
	\end{tabular}
	\begin{tablenotes}
	    \footnotesize
        \item \textbf{Notes.}
        \item[a] Right ascension.
        \item[b] Declination.
        \item[c] \textcolor{black}{Best-fit redshift, from principal component analysis of SDSS galaxy eigenspectra from BOSS. G11/G18 have only one spectral feature.}
        \item[d] Apparent HST ACS+F814W magnitude.
        \item[e] Absolute \textit{B-}band magnitude.
        \item[f] \textcolor{black}{Stellar mass from stellar population fits to the MUSE spectrum and DES \& HST photometry.}
        \item[g] Angular distance from the quasar.
        \item[h] Projected physical distance from the quasar. 
        \item[i] LOS velocity from the quasar.  
        \item[j] Stellar mass estimated from the median $M_*/L$ ratio of the group resulting in large systematic uncertainties.  
        \item[k] The uncertainty in the position of G3 is larger than other galaxies due to the diffraction spike in the HST ACS+F814W image.  
    \end{tablenotes}
	\end{threeparttable}
\end{table*}

To examine properties of the quasar host environment, we identified candidate group members based on their LOS velocities relative to the quasar ($\Delta v = v - v_{\rm QSO}$). In particular, we selected galaxies with $|\Delta v| < \rm 2000\ km \, s^{-1}$. We inferred the physical properties of the selected galaxies with \texttt{Bagpipes} \citep{2018MNRAS.480.4379C, 2019MNRAS.490..417C}. \texttt{Bagpipes} performs stellar population synthesis (SPS) with a stellar evolution model from \citet{2003MNRAS.344.1000B}, an initial mass function from \citet{2001MNRAS.322..231K}, and the Bayesian inference package \texttt{Multinest} \citep{2014A&A...564A.125B, 2009MNRAS.398.1601F, 2019OJAp....2E..10F}. \textcolor{black}{We fit both spectroscopic and photometric data simultaneously with \texttt{Bagpipes}.} Many of the galaxies in our sample only have one photometric datapoint available, necessitating the use of the spectra to further inform the stellar population synthesis. In our fitting procedure, we assumed an exponential star formation history with e-folding time scale of $0.01 < \rm \tau/Gyr <8.00$, solar stellar metallicity, and dust attenuation model from \citet{2000ApJ...533..682C} with $0 < A_V/\rm mag < 2$. \textcolor{black}{The choice of exponentially declining star formation histories enables more direct comparison with surveys such as MUSE-Wide \citep{2019A&A...624A.141U} and the MUSE Ultra DEEP Field \citep{2019MNRAS.490.1451F}}. \textcolor{black}{We introduced a 2nd order multiplicative polynomial to reconcile the potential artificial differences between SED measured in photometry and spectra. This polynomial accounts for systematic uncertainty in the MUSE flux due to wavelength dependent aperture losses and uncertainty in the flux calibration \citep{2020A&A...641A..28W}. We also used \texttt{Bagpipes} spectrum noise scaling to allow the relative weighting of the photometry and spectrum to be a nuisance parameter. We note that the results are not sensitive to this scaling in our case \citep[see][]{2019MNRAS.490..417C}.} In addition to the ACS$+$F814W photometry, we also included $grizY$ photometric data from the Dark Energy Survey (DES; \citealt{2021ApJS..255...20A}) available for 16 galaxies. The resulting stellar mass estimates \textcolor{black}{and dust attenuation $A_V$ values} are reported in Table \ref{tab:summary_galaxies}. The stellar masses have associated systematic uncertainties of $\approx0.2$ dex. Galaxies close to the quasar (G1-G7) are contaminated by the quasar light, and we used the quasar-light subtracted spectra for \texttt{Bagpipes} fitting when possible. Galaxies G1, G3, G11, G13, G18, G20, and G31 do not have a stellar mass estimate because their continua are too faint or are too badly contaminated by the quasar continuum. To further characterize these galaxies, we also report $4000$\,\AA \, break strength (D4000; \citealt{2005MNRAS.362...41G}) and rest-frame $B$-band absolute magnitude with $K$-corrections calculated using templates from \citet{1980ApJS...43..393C} chosen based on \textcolor{black}{the strength of the 4000 \AA\ break}. The IDs, galaxy coordinates (R.A., Decl.), redshifts, ACS$+$F814W apparent magnitudes, absolute $B$-band magnitudes, adopted K-correction templates (S0, Scd, or Irregular), and D4000 measurements are reported in Table \ref{tab:summary_galaxies}, along with the angular distances, projected distances, and LOS velocity differences from the quasar sightline. The locations of these galaxies are shown in Figure \ref{fig:dis} and \textcolor{black}{several example} MUSE spectra are overplotted with their best-fit PCA spectral models in Figure \ref{fig:gal}. \textcolor{black}{An interactive view of the galaxy environment and spectra is available online\footnote{\label{www}\url{http://zhuoqiliu.com/HE0238-1904.html}}}.

\subsection{The Galactic Environment}
\label{GE}
In the MUSE field of HE\,0238$-$1904 we identified 35 galaxies, including the quasar host, with LOS velocities $|\Delta v| < \rm 2000 \, km \, s^{-1}$ of the quasar systemic velocity, which is sufficient to encompass most members of even massive galaxy clusters. Figure \ref{fig:dis} shows a $1.5'\times 1.5'$ FoV image from the ACS+F814W observations of the field where we marked the quasar with a grey star and labelled galaxies with circles as well as their ID. The color of the circle represents the LOS velocity of each galaxy relative to the quasar. Additionally, we display the $1'\times 1'$ MUSE FoV, and a smaller $30'' \times 30''$ region which is the focus of later figures in this work.

Among the 35 galaxies in the environment of HE\,0238$-$1904, \textcolor{black}{four (two) exhibit stellar masses of $\log(M_*/{\rm M_{\odot}})>10.5\ (> \!11)$ (excluding the quasar)}, indicating a significant overdensity and likely a massive group. To further characterize the environment, we show the distribution of galaxies' LOS velocities relative to the quasar ($\Delta v = v - v_{\rm QSO}$) in the bottom right panel of Figure \ref{fig:dis}. The LOS velocity distribution peaks around $\rm -100 \, km \, s^{-1}$ but exhibits a non-Gaussian tail toward higher velocity of $\rm +100\, km \, s^{-1}$ to $\rm +1400 \, km \, s^{-1}$. There is a clear trend between LOS velocity and location on the sky visible in Figure \ref{fig:dis} with galaxies with $\Delta v > 0 \rm \, km \, s^{-1} $ largely falling North East of the quasar and those with $\Delta v < 0 \, \rm km \, s^{-1} $ falling near the quasar or South West of it. To better visualize the location$-$velocity trend, we divided the field into two regions, \textcolor{black}{one NE of the quasar and one SW of it. The NE (SW) }one is marked by an orange (purple) trapezoid in Figure \ref{fig:dis}. We also show the LOS velocity distribution of the galaxies in each trapezoidal region by the corresponding histograms in the inset panel in Figure \ref{fig:dis}. The peak and the tail in the histogram correspond closely to these two regions respectively. The non-Gaussian LOS velocity distribution and correlation with spatial location suggests that the overdensity near the quasar host may consist of two distinct, but possibly interacting, galaxy groups.

To quantify the velocity dispersions of these two potential groups, we fit two Gaussians to the entire LOS velocity distribution. This results in one narrow, blueshifted Gaussian and one broader, redshifted one. The blueshifted Gaussian has a mean LOS velocity of $\Delta v_{\rm group} = \rm -99 \pm 25 \, km \, s^{-1}$ and a 1D velocity dispersion of $\sigma_{\rm group} = \rm 92 \pm 50 \, km \, s^{-1}$ and includes $\approx 35\%$ of the galaxies near HE\,0238$-$1904. The redshifted Gaussian has $\Delta v_{\rm group} = \rm 629 \pm 140 \, km \, s^{-1}$ and $\sigma_{\rm group} = \rm 506 \pm 90 \, km \, s^{-1}$ and includes $\approx 65\%$ of the galaxies. In both cases, the uncertainty estimates are based on bootstrap resampling. While the Gaussian fitting did not include any spatial information, the two Gaussians closely match the purple and orange velocity histograms formed from a spatial separation (see Figure \ref{fig:dis}). These fitting results suggest that the environment around the quasar includes one massive group at $\Delta v_{\rm group} \approx \rm 600 \, km \, s^{-1}$ and one less massive group closer to the quasar velocity. Assuming each group is virialized, we estimate dynamical masses of $M_{\rm dyn} \sim 9.8 \times 10^{13}\ {\rm M_{\odot}}$ and $M_{\rm dyn}\sim 5.7 \times 10^{11}\ {\rm M_{\odot}}$ \citep{2013MNRAS.430.2638M} for the richer, redshifted group and less rich, blueshifted group, respectively. \textcolor{black}{To place a lower limit on the mass estimate, we fit a single Gaussian to galaxies with $\Delta v > 200 \, \rm km \, s^{-1}$. We found the velocity dispersion is $\approx 400 \, \rm km \, s^{-1}$, corresponding to a mass of $M_{\rm dyn} \sim 3.8 \times 10^{13}\ {\rm M_{\odot}}$. The mass range of $M_{\rm dyn} \approx 4 \times 10^{13}-10^{14}\ {\rm M_{\odot}}$ is consistent with massive group or modest mass cluster.} \textcolor{black}{However, we caution that the assumption that the groups are virialized introduces additional uncertainty given the complex environment}. \textcolor{black}{Finally, in Figure \ref{fig:dis}, we show the stellar mass weighted group center as a white asterisk, and membership weighted ($\frac{P_{\rm blue/red}}{P_{\rm blue} + P_{\rm red}}$) centers as red and blue asterisks for the richer, redshifted group and less rich, blueshifted group respectively.}

To test the expectation that dynamically more massive groups will contain more massive galaxies, we investigate the most massive galaxies in each group. G8 and G22 are the most massive galaxies with a stellar mass of $\log(M_*/{\rm M_{\odot}})=10.4$ and $10.1$ respectively in the less rich, blueshifted group. On the other hand, the richer, redshifted group includes two massive elliptical galaxies, G10 and G34, with $\log(M_*/\mathrm{M}_{\odot})=11.5$ and $11.2$, respectively. Furthermore, the richer, redshifted group contains a massive disc galaxy, G33, with $\log(M_*/{\rm M_{\odot}})=10.8$. This is consistent with HE\,0238$-$1904 residing in an overdense region likely made of two groups with the redshifted one being richer and more massive. However, the quasar redshift falls between the centroids of the two groups indicating that it could arise in either or truly be located between them. \textcolor{black}{Despite the large uncertainty in the stellar mass of the quasar host galaxy (see Section \ref{MEQP}), the large black hole mass suggests it is a massive galaxy, possibly the largest in the overdensity around HE\,0238$-$1904. It is therefore more probable that HE\,0238$-$1904 resides in the richer, redshifted group. Nonetheless, we cannot completely rule out the possibility that HE\,0238$-$1904 originates from the less rich, blueshifted group.} In either case, the dynamically rich and likely unrelaxed environment could result in galaxy interactions that can produce giant nebulae via ram pressure and tidal stripping. 

\begin{figure*}
    \centering
    \includegraphics[scale=0.35, align=t]{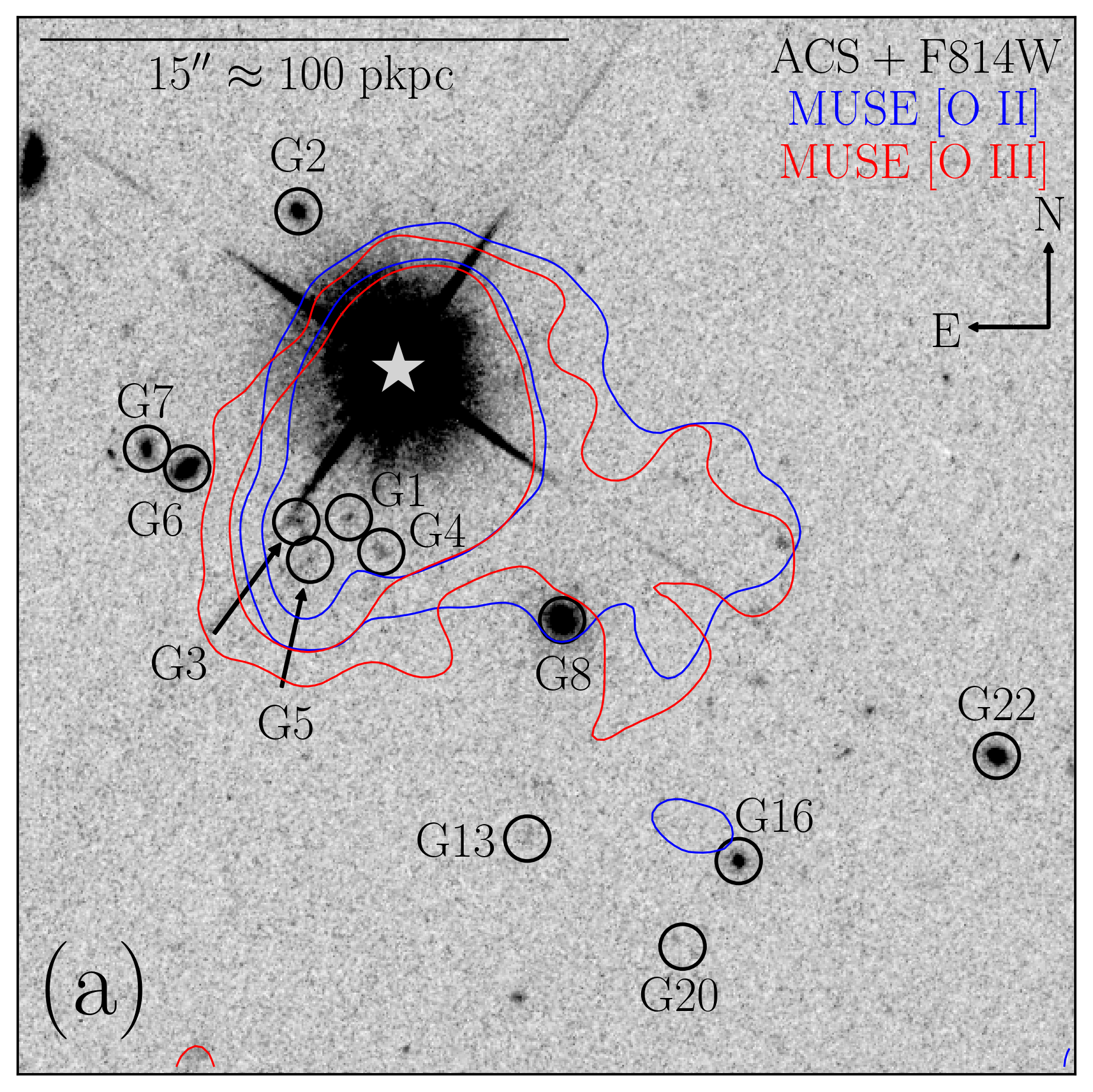}
    \includegraphics[scale=0.35, align=t]{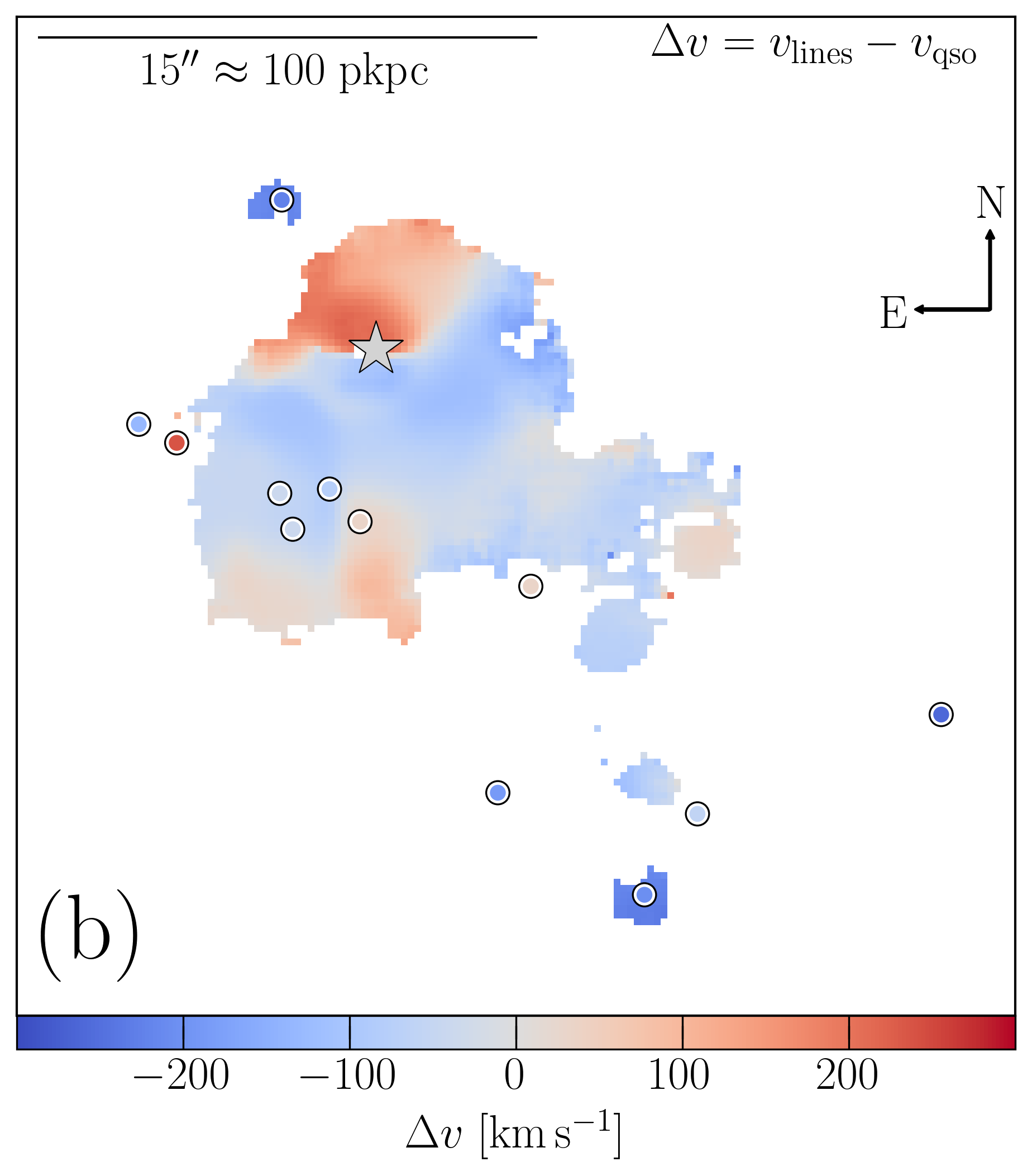}
    \includegraphics[scale=0.35, align=t]{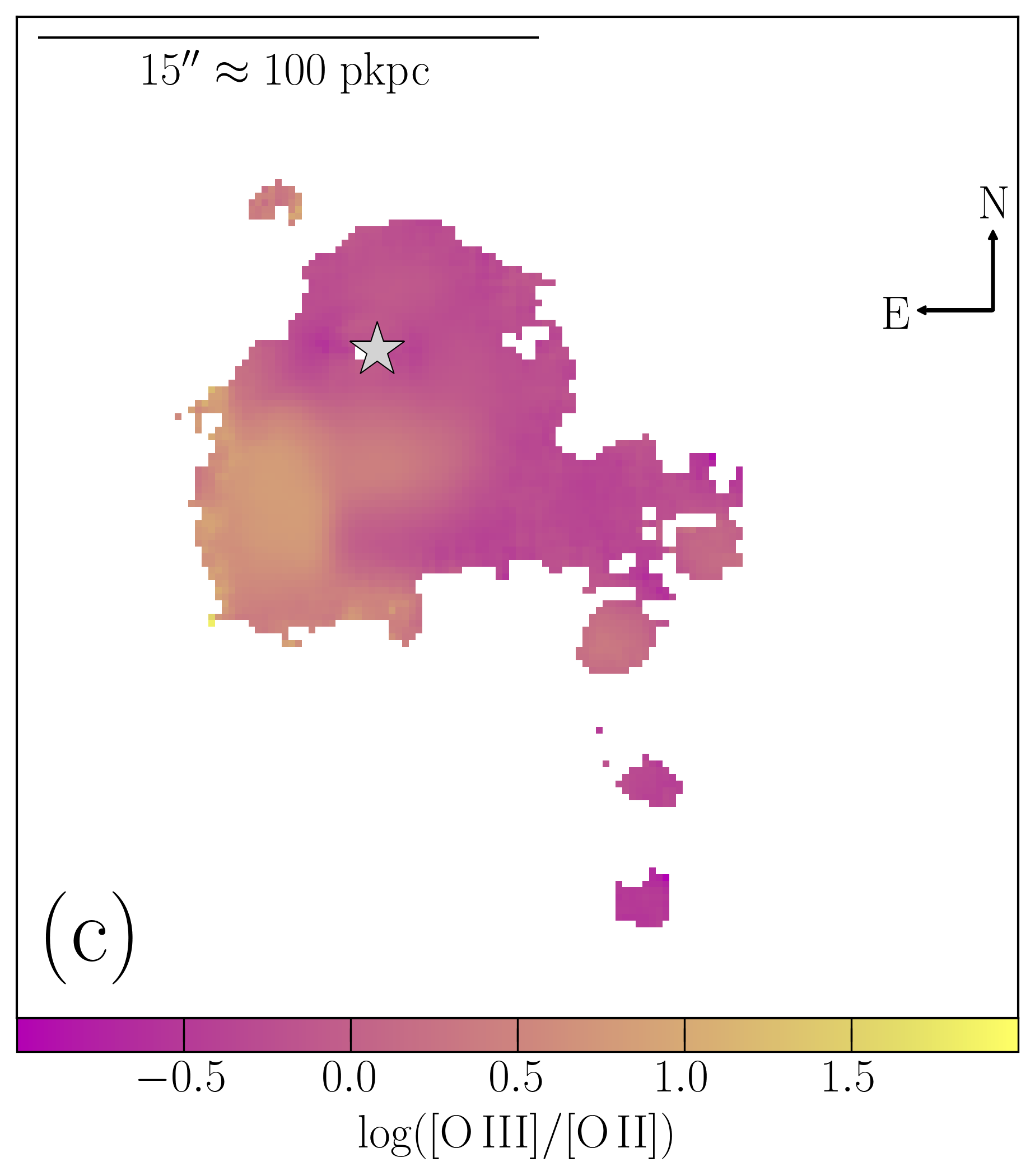}
    \includegraphics[scale=0.35, align=c]{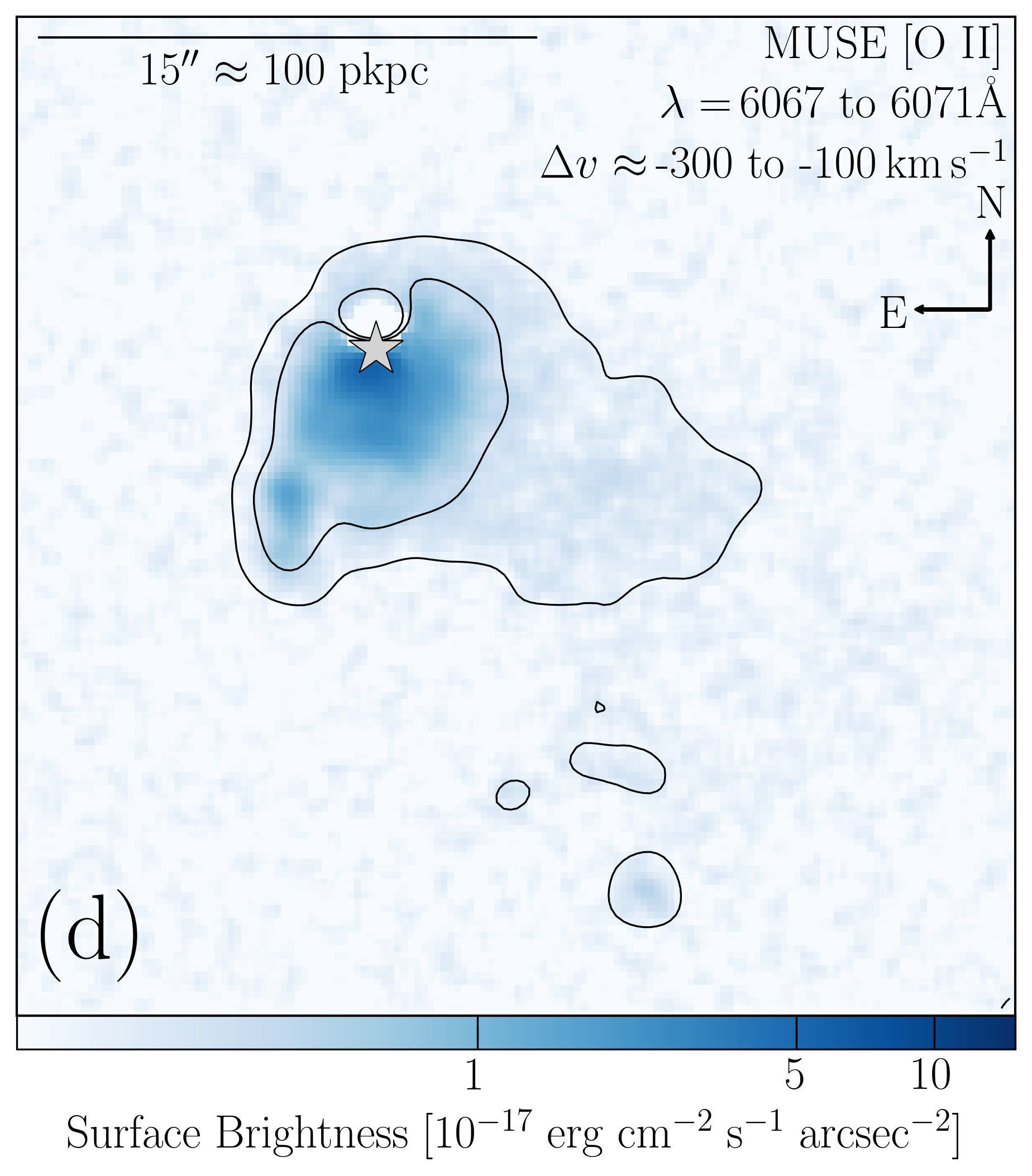}
    \includegraphics[scale=0.35, align=c]{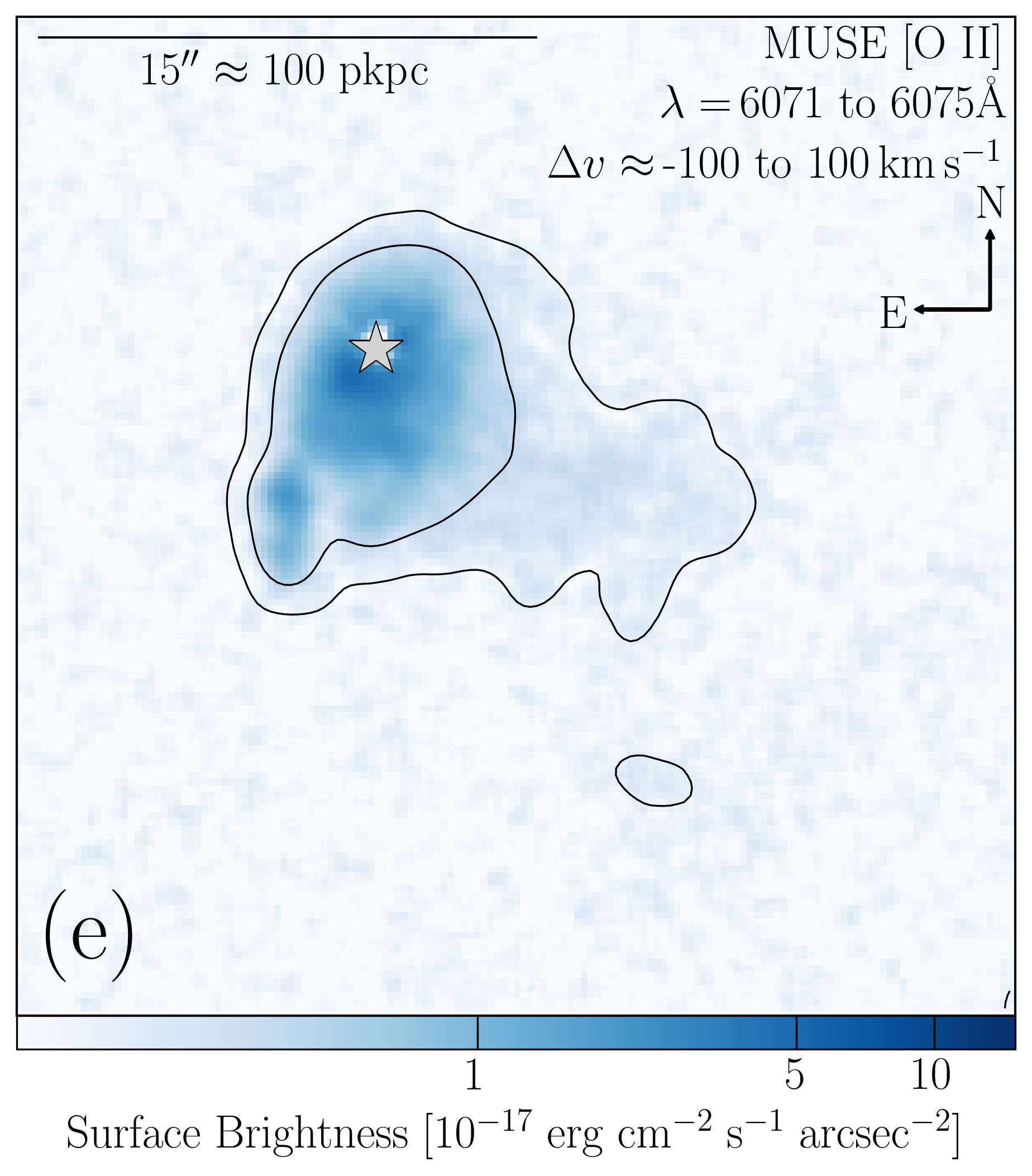}
    \includegraphics[scale=0.35, align=c]{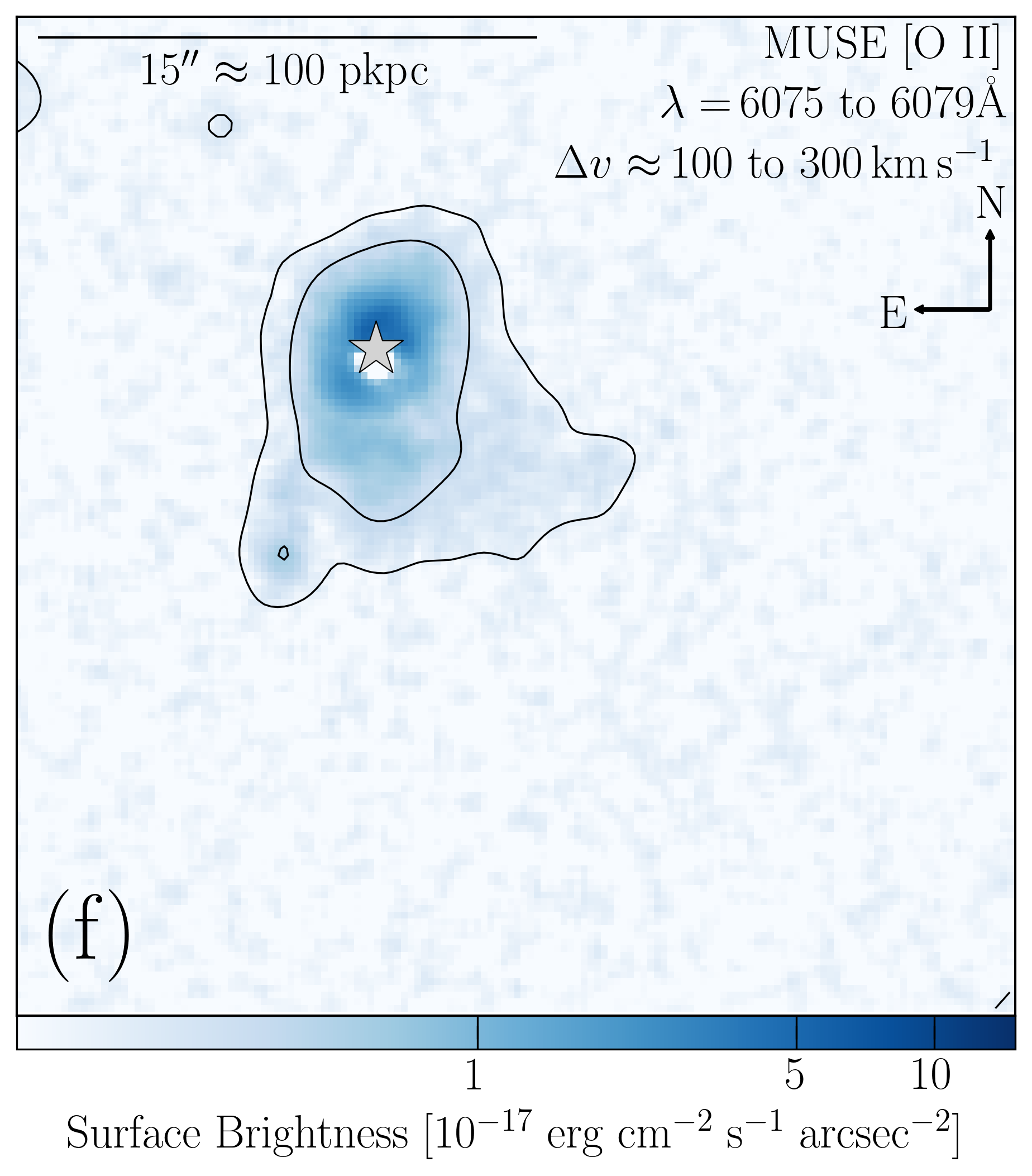}
    \includegraphics[scale=0.35, align=c]{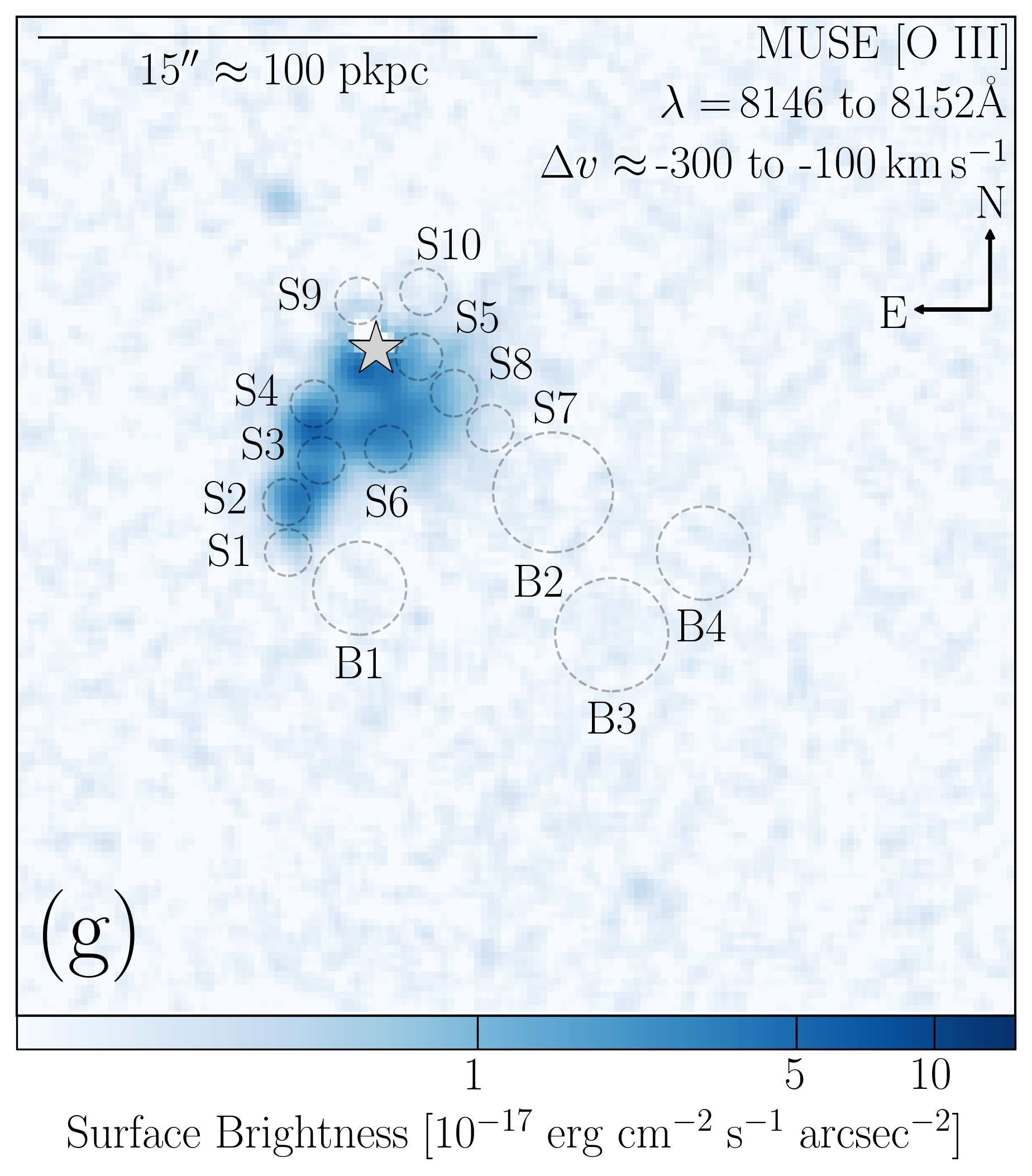}
    \includegraphics[scale=0.35, align=c]{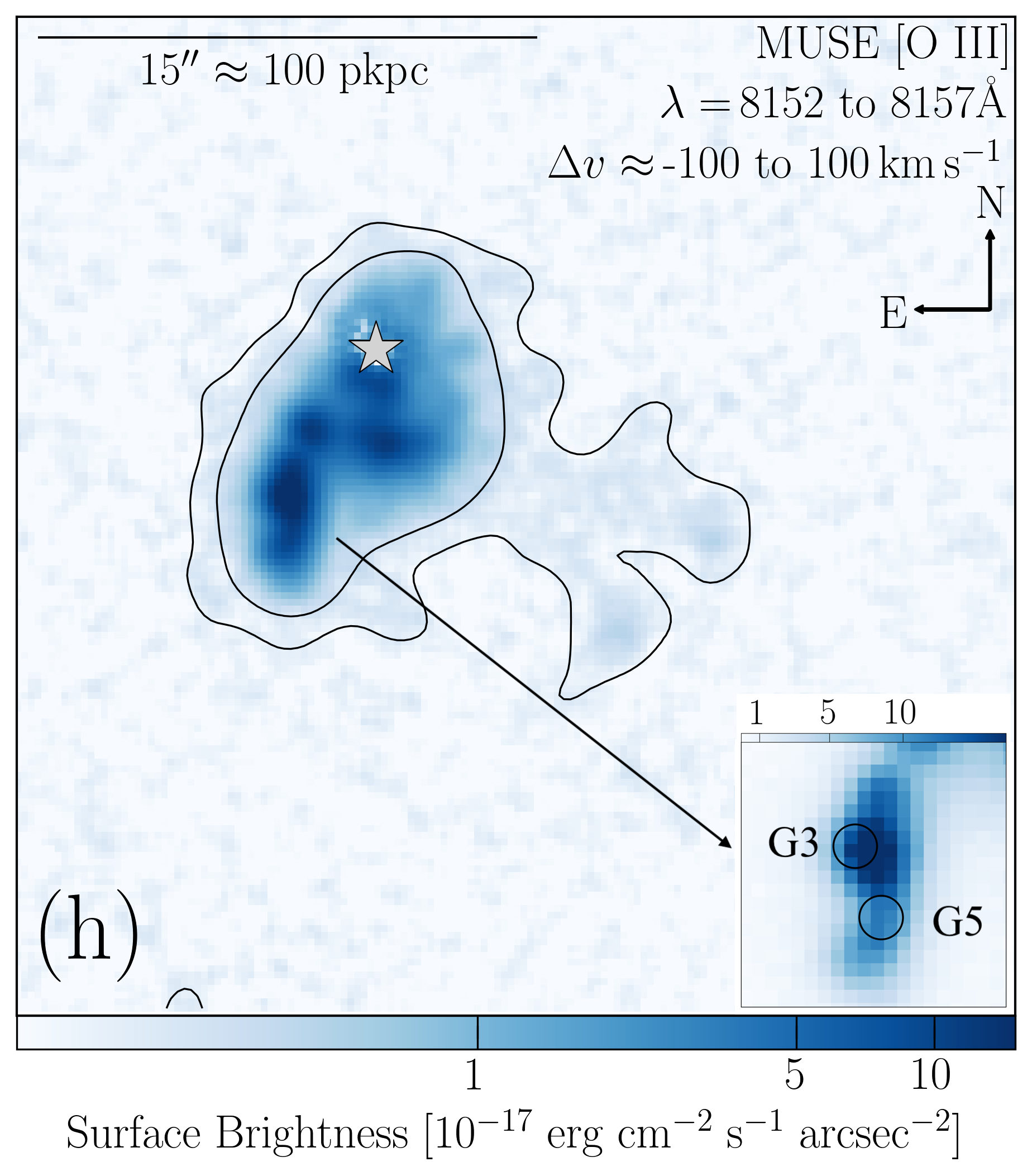}
    \includegraphics[scale=0.35, align=c]{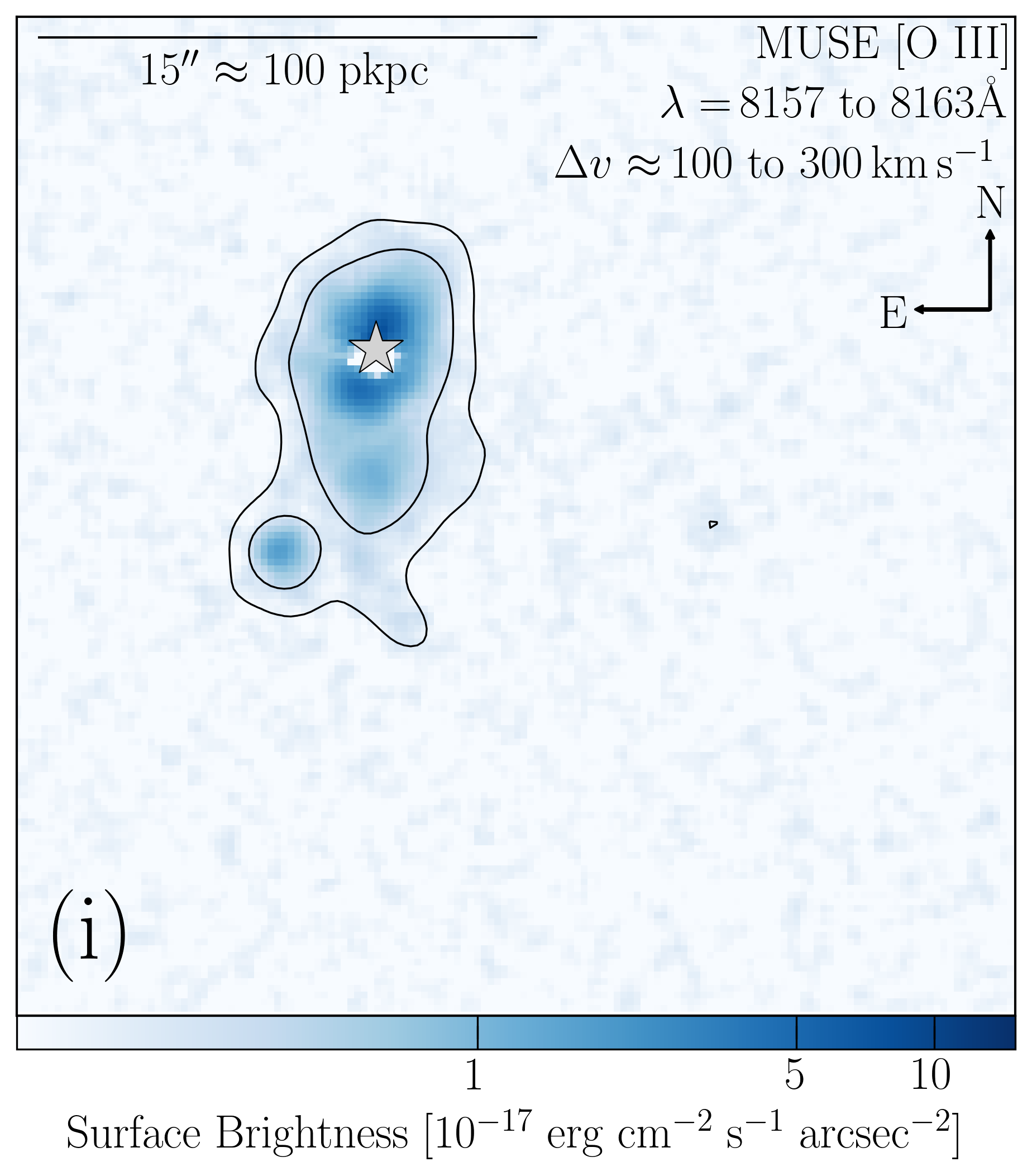}
    \caption{Visualizations of the nebula discovered around HE\,0238$-$1904. Panel (a): HST ACS+F814W image of the field. Galaxies are circled in black and labelled with their IDs.
    Panel (b): map of the nebular LOS velocity relative to the quasar systemic velocity. Galaxies are circled in black and colored with their velocities. Panel (c): map of nebular photoionization shown as the line ratio $\rm [O \, III]\lambda5008 / [O \, II]\lambda\lambda3727+3729$. 
    Panel(d)-(f) and Panel (g)-(i): narrow-band $\rm [O \, II]$ and $\rm [O \, III]$ surface brightness maps extracted from the MUSE datacube over the velocity intervals labelled in each panel. \textcolor{black}{The inset panel in Panel (h) shows a zoomed, unsmoothed map around G3 and G5 to emphasize the possible existence of a tidal tail}. These maps are overlaid with $\rm [O \, II]$ and $\rm [O \, III]$ surface brightness contours at levels of $0.08$ and $0.3 \times 10^{-17} \rm \, erg \, cm^{-2} \, s^{-1} \, arcsec^{-2}$. The contours shown in panel (e) and panel (h) are overlaid on the HST image in blue and red respectively. We note that surface brightness maps and contours are smoothed with $3$ pixel kernels. \textcolor{black}{A version of this figure with the region circles marked in every velocity panel is available online\footref{www}}.} 
    \label{fig:gas}
\end{figure*}

\begin{table*}
	\centering
	\caption{Summary of emission-line measurements for extracted regions in the nebula around HE\,0238$-$1904.}
	\label{tab:gas_pc}
	\begin{threeparttable}
	\begin{tabular}{lrccccccccc} 
		\hline
		ID & Distance\tnote{a} & Extraction & $\rm [O \, II]$ & $\rm H\beta$ & $\rm[O \, III]$ & $\rm[Ne \, V]$ & $\rm[O \, III]$ & $\rm He \, II$ & \multicolumn{1}{c}{$\Delta$\textit{v}\tnote{b}} & \multicolumn{1}{c}{\textit{$\sigma$}\tnote{c}} \\
		& ($\rm pkpc$) & radius & $\lambda\lambda3727+3729$ & & $\lambda5008$ & $\lambda3346$ & $\lambda4364$ & $\lambda4687$ & ($\mathrm{km\ s^{-1}}$) & ($\mathrm{km\ s^{-1}})$\\
        & & ($''$) & ($\mathrm{10^{-17} \, erg}$ & ($\mathrm{10^{-17} \, erg}$ & ($\mathrm{10^{-17} \, erg}$ & ($\mathrm{10^{-17} \, erg}$ & ($\mathrm{10^{-17} \, erg}$ & ($\mathrm{10^{-17} \, erg}$ & & \\
  		& & & $\mathrm{s^{-1} cm^{-2}}$) & $\mathrm{s^{-1} cm^{-2}}$) & $\mathrm{s^{-1} cm^{-2}}$) & $\mathrm{s^{-1} cm^{-2}}$) & $\mathrm{s^{-1} cm^{-2}}$) & $\mathrm{s^{-1} cm^{-2}}$) & &\\
		\hline
		S1 & 45 & 0.7 & $1.73 \pm 0.05$ & $0.69 \pm 0.06$ & $9.17 \pm 0.05$ & $0.15 \pm 0.03$ & $0.21 \pm 0.02$ & $\hspace{-11.0mm}<0.21$ & $-11 \pm 3$ &  $62 \pm 4$ \\
		S2 & 36 & 0.7 & $3.55 \pm 0.08$ & $1.14 \pm 0.14$ & $\hspace{-1.5mm} 23.48 \pm 0.10$ & $0.37 \pm 0.05$ & $0.40 \pm 0.04$ & $0.35 \pm 0.11$ & 
        $-55 \pm 3$ & $43 \pm 4$ \\
		S3 & 25 & 0.7 & $\hspace{-11.0mm}<0.30$  & $\hspace{-11.0mm}<0.27$ & $6.27 \pm 0.22$ & $\hspace{-11.0mm}<0.15$ & $\hspace{-11.0mm}< 0.09$ & $\hspace{-11.0mm}<0.18$ & $\hspace{-1.5mm}-107 \pm 3$ & $61 \pm 4$ \\
  		$\rm S3_{wing}$ & 25 & 0.7 & $2.90 \pm 0.10$ & $0.73 \pm 0.09$ & $2.44 \pm 0.22$ & $\hspace{-11.0mm}< 0.18$ & $\hspace{-11.0mm} < 0.12$ & $\hspace{-11.0mm} < 0.21$ & $-14 \pm 9$ & $\hspace{-1.5mm}104 \pm 5$ \\
		S4 & 17 & 0.7 & $1.34 \pm 0.18$ & $0.28 \pm 0.08$ & $3.39 \pm 0.10$ & $\hspace{-11.0mm}<0.18$ & $\hspace{-11.0mm}<0.09$ & $\hspace{-11.0mm}<0.15$ & $\hspace{-1.5mm}-114 \pm 3$ & $45 \pm 4$ \\
  		$\rm S4_{wing}$ & 17 & 0.7 & $4.17 \pm 0.20$ & $0.52 \pm 0.09$ & $3.14 \pm 0.12$ & $\hspace{-11.0mm} < 0.27$ & $\hspace{-11.0mm} < 0.15$ & $\hspace{-11.0mm} < 0.27$ & $\hspace{0.3mm}+12 \pm 8$ & $\hspace{-1.5mm}169 \pm 6$ \\
		S5 & 9 & 0.7 & $5.96 \pm 0.28$ & $0.77 \pm 0.26$ & $2.51 \pm 0.22$ & $\hspace{-11.0mm}< 0.84$ & $\hspace{-11.0mm}< 0.51$ & $\hspace{-11.0mm}< 0.78$ & $\hspace{3.1mm}+8 \pm 11$ & $\hspace{0.0mm}140 \pm 11$ \\
		S6 & 20 & 0.7 & $5.04 \pm 0.07$ & $1.47 \pm 0.12$ & $\hspace{-1.5mm} 14.03 \pm 0.07$ & $0.15 \pm 0.05$ & $0.22 \pm 0.04$ & $0.34 \pm 0.09$ & $-62 \pm 3$ & $68 \pm 4$\\
		S7 & 29 & 0.7 & $0.99 \pm 0.04$ & $0.18 \pm 0.06$ & $0.63 \pm 0.04$ & $\hspace{-11.0mm} < 0.09$ & $\hspace{-11.0mm}< 0.06$ & $\hspace{-11.0mm}< 0.18$ & $-72 \pm 8$ & $\hspace{-1.5mm}111 \pm 8$\\
		S8 & 18 & 0.7 & $2.33 \pm 0.04$ & $0.52 \pm 0.06$ & $1.98 \pm 0.04$ & $\hspace{-11.0mm}< 0.09$ & $\hspace{-11.0mm}< 0.06$ & $\hspace{-11.0mm}< 0.15$ & $\hspace{-1.5mm}-119 \pm 4$ & $89 \pm 4$ \\
		S9 & 11 & 0.7 & $3.71 \pm 0.16$ & $1.10 \pm 0.15$ & $2.56 \pm 0.13$ & $\hspace{-11.0mm}< 0.45$ & $\hspace{-11.0mm}< 0.27$ & $\hspace{-11.0mm} < 0.39$ & $\hspace{-1.0mm}+173 \pm 7$ & $\hspace{-1.5mm}110 \pm 7$ \\
		S10 & 15 & 0.7 & $1.96 \pm 0.05$ & $0.47 \pm 0.05$ & $1.58 \pm 0.04$ & $\hspace{-11.0mm} < 0.12$ & $\hspace{-11.0mm} < 0.09$ & $\hspace{-11.0mm} < 0.15$ & $\hspace{0.3mm}+58 \pm 4$ & $79 \pm 5$ \\
		B1 & 49 & 1.4 & $1.14 \pm 0.08$ & $0.89 \pm 0.12$ & $2.21 \pm 0.08$ & $\hspace{-11.0mm} < 0.21$ & $\hspace{-11.0mm} < 0.15$ & $\hspace{-11.0mm} < 0.33$ & $\hspace{0.3mm}+50 \pm 6$ & $\hspace{-1.5mm}128 \pm 7$ \\
		B2 & 47 & 1.8 & $4.32 \pm 0.13$ & $\hspace{-11.0mm} < 0.57$ & $1.96 \pm 0.11$ & $\hspace{-11.0mm} < 0.30$ & $\hspace{-11.0mm} < 0.21$ & $\hspace{-11.0mm} < 0.57$ & $-36 \pm 8$ & $\hspace{-1.5mm}119 \pm 8$ \\
  		B3 & 76 & 1.7 & $1.03 \pm 0.09$ & $\hspace{-11.0mm} < 0.60$ & $1.37 \pm 0.07$ & $\hspace{-11.0mm} < 0.21$ & $\hspace{-11.0mm} < 0.15$ & $\hspace{-11.0mm} < 0.42$ & $-69 \pm 5$ & $50 \pm 6$ \\
        B4 & 79 & 1.4 & $0.31 \pm 0.11$ & $\hspace{-11.0mm} < 0.24$ & $0.83 \pm 0.06$ & $\hspace{-11.0mm} < 0.18$ & $\hspace{-11.0mm} < 0.12$ & $\hspace{-11.0mm} < 0.33$ & $\hspace{0.3mm}+30 \pm 4$ & $30 \pm 6$ \\ 
        $\rm B4_{wing}$ & 79 & 1.4 & $0.99 \pm 0.16$ & $\hspace{-11.0mm} < 0.24$ & $0.40 \pm 0.11$ & $\hspace{-11.0mm} < 0.18$ & $\hspace{-11.0mm} < 0.12$ & $\hspace{-11.0mm} < 0.33$ & $\hspace{1.3mm}-83 \pm 42$ & $201 \pm 36$\\

		\hline \\
	\end{tabular}
	\begin{tablenotes}
	    \footnotesize
        \item[a] Projected physical distance from the quasar. 
        \item[b] LOS velocity relative to the quasar with uncertainty, assuming a systematic uncertainty of $3 \rm \, km \, s^{-1}$ \citep{2020A&A...641A..28W}.
        \item[c] LOS velocity dispersion with uncertainty, assuming a systematic uncertainty of $4 \rm \, km \, s^{-1}$ \citep{2016A&A...588A.149K}.  
    \end{tablenotes}
	\end{threeparttable}
\end{table*}

\subsection{Nebular Environment}
\label{NE}
Due to ionizing radiation from the accretion disk, wide-field IFS observations of quasar fields often find large  nebulae (Johnson et al., in prep). To search for the nebulae around HE\,0238$-$1904, we conducted continuum subtraction of the datacube locally for the $\rm [O \, II]$, $\rm H\beta$, and $\rm [O \, III]$ emission lines around the quasar. For continuum fitting near each of the three lines, we masked the spectral region within $\pm 500{-}1000 \, \rm km\, s^{-1}$ of the expected observed wavelength at the quasar's redshift. We fine-tuned the masked region individually for each of the three lines to avoid skyline contamination and to account for the larger width $\rm [O \, II]$ doublet. For each spaxel in the masked datacube, we then fit a third-order polynomial to the continuum regions around each line and subtracted the best-fit model to complete the continuum subtraction.

This continuum-subtracted MUSE datacube enabled the discovery of a giant ionized nebula in $\rm [O \, II]$, $\rm H\beta$, and $\rm [O \, III]$ around HE\,0238$-$1904 with a total area of $\approx 5000\ {\rm kpc}^2$ which is visualized in Figure \ref{fig:gas}. This nebula surrounds the quasar with projected radii of $d \rm \approx \! 30 \ to \ 50 \, pkpc$ and with LOS velocities of $\Delta v \approx -250 \ \rm to \ {+}250 \ km \, s^{-1}$ from the quasar. The nebula is more extended to the South East and the South West of the quasar. The South East extension of the nebula is spatially coincident with galaxies G1, G3, G4, and G5. Additionally, the tail extending South West of the quasar is distinct from but approximately in the direction of G8.

To examine the nebula and any relationship with galaxies in the quasar environment, we show $\rm [O \, II]$ and $\rm [O \, III]$ emission contours over the HST image in panel (a) of Figure \ref{fig:gas}. We also display a nebular LOS velocity map in panel (b) and a $\rm [O \, III]/[O \, II]$ line ratio map in panel (c). We constructed these two maps by jointly fitting  Gaussian line profiles to the continuum-subtracted $\rm [O \, II]$, $\rm H\beta$, and $\rm [O \, III]$ datacubes. Instead of fitting the spectrum of each individual spaxel, we averaged over circular apertures of $r=1''$ to enhance S/N. We chose this aperture radius based on experimentation to visualize even faint parts of the nebula. These two maps provide an opportunity to study the spatial dependence of the kinematics and the ionization state of the gas. In addition, we show three panels of narrowband images generated from the continuum subtracted datacubes for each of $\rm [O \, II]$ and $\rm [O \, III]$ in velocity ranges of $-300 \rm \ to \ {-}100 \ km \, s^{-1}$, $-100 \rm \ to \ {+}100 \ km \, s^{-1}$, and $+100 \rm \ to\ {+}300 \ km \, s^{-1}$ in \textcolor{black}{panel (d)-(f) and (g)-(i) respectively.}

The nebula exhibits an irregular morphology but with a spatial trend in kinematics. In particular, the region North of the quasar is redshifted relative to the quasar and has a LOS velocity of $\Delta v = 0{-}250\rm \, km \, s^{-1}$. The region South of the quasar including the tail to the West is mainly blueshifted relative to the quasar but with a small redshifted region in the most Southern points. This southern region is spatially coincident and potentially kinematically coincident with G1, G3, G4 and G5. \textcolor{black}{However, the continua of these galaxies are too faint to measure stellar absorption-based redshifts. This raises the possibility that their nebular spectra may be contaminated by the surrounding nebulae, resulting in a biased redshift measurement. In the case of G3 and G4, the line width of the nebular emission near the galaxies is significantly narrower than  the more extended emission from nearby parts of the giant nebula, indicating that the galaxy line emission likely arises in the ISM of the two dwarfs}.

The nebula also shows a spatial trend in the ionization-state-sensitive $\rm [O \, III] / [O \, II]$ line ratio. The majority of the nebula is $\rm [O \, II]$ dominated but the region South East of the quasar has greater $\rm [O \, III]$ emission, particularly, at a few $\rm [O \, III]$ knots near G1, G3 and G5. The knots near G3 and G5 have the highest surface brightness in the nebula. \textcolor{black}{Furthermore, the bright region extending to the South of the brightest knot near G3 is reminiscent of a} tidal tail.

To better explore the properties of the nebula, we selected several representative regions in it and extracted their full spectra to infer physical conditions from both strong ($\rm [O \, II]$, $\rm H\beta$, and $\rm [O \, III]$) and weak lines ($\rm [Ne \, V]\lambda3427$, $\rm H\delta$, $\rm H\gamma$, $\rm [O \, III]\lambda4364$, and $\rm He \,II \lambda4687$\footnote{Other weak lines such as $\rm [Ne \,III]\lambda3869$, $\rm He \,I \lambda 3889$ \& $\rm H8$, and $\rm H\epsilon$ are covered by MUSE but we do not use them in this work because of contaminating sky lines or blending with other lines.}). We picked the locations of these regions to cover a wide range in line ratios, surface brightness, and projected locations relative to the quasar. These regions are shown in panel (g) of Figure \ref{fig:gas} and labelled with letters and numbers where S\# refers to regions with higher surface brightness for which we used an extraction radius of $0.7''$ while B\# labels low surface brightness regions which required a larger extraction radius ($>1''$) to achieve sufficient S/N.

\begin{figure*}
    \centering
    \includegraphics[scale=0.7]{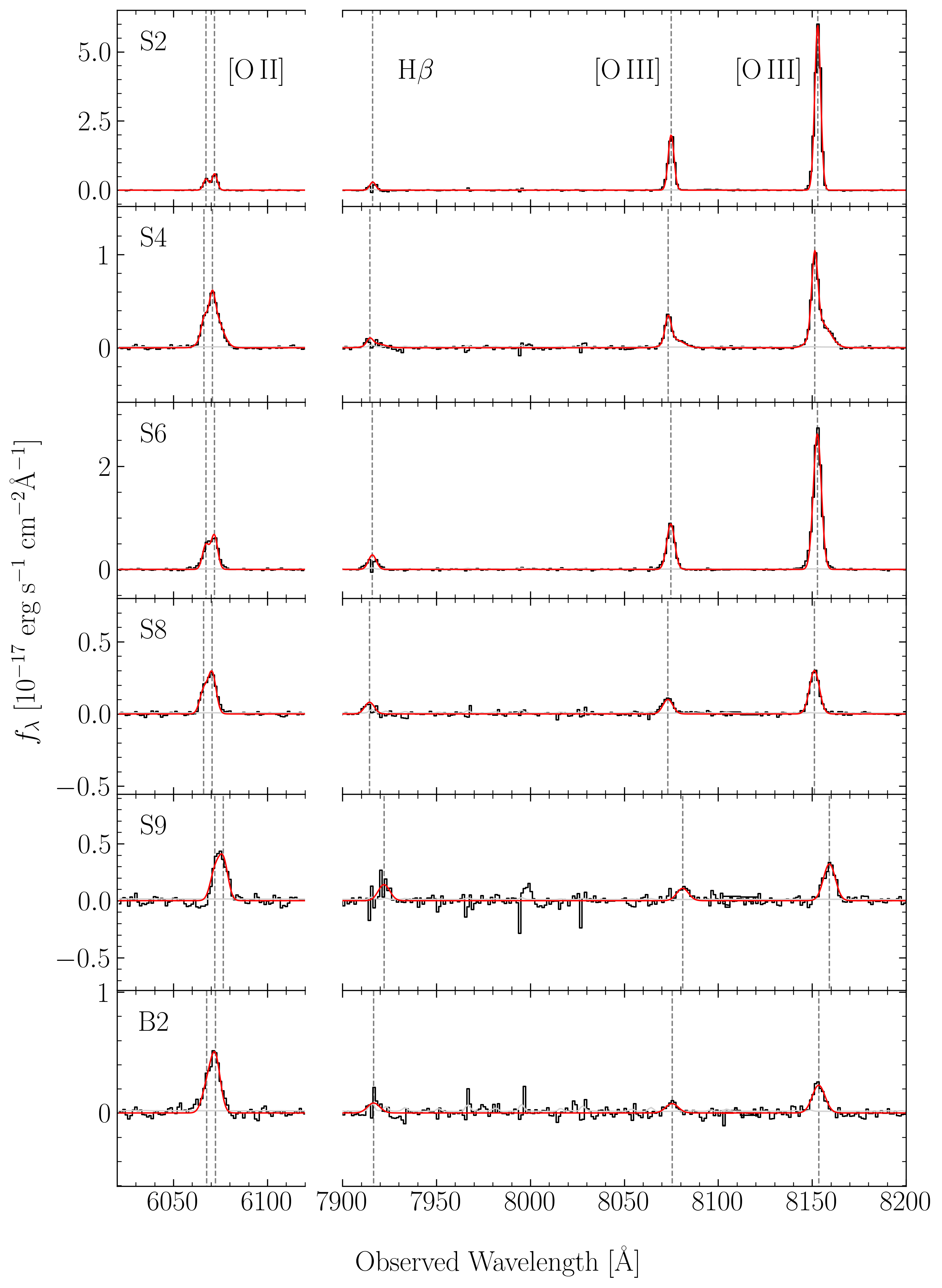}
    \caption{Examples of nebular spectra (stronger lines) and best-fit spectral models for multiple regions. The locations of these regions are shown as circles and labelled by their IDs in Figure \ref{fig:gas}. The extracted spectrum is shown as solid black lines and the error array is shown as grey lines. The best-fit models are shown as red solid lines. In most nebular regions, we detected strong emission lines such as $\rm [O \, II]$, $\rm H\beta$, and $\rm [O \, III]$.}
    \label{fig:gas_strong}
\end{figure*}

\begin{figure*}
    \centering
    \includegraphics[scale=0.7]{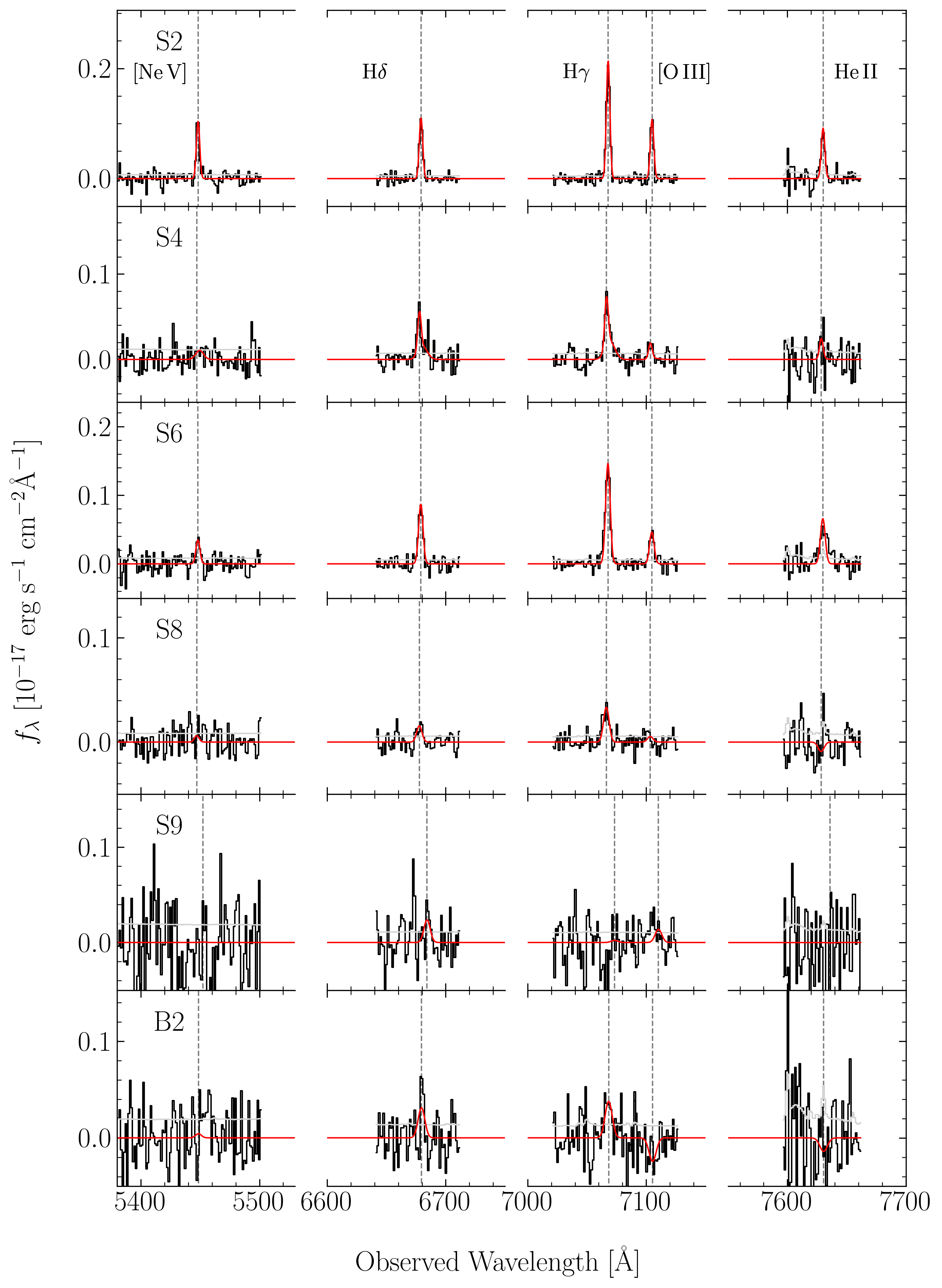}
    \caption{Examples of nebular spectra (fainter lines) and best-fit spectral models for multiple regions. The locations of these regions are shown as circles and labelled by their IDs in Figure \ref{fig:gas}. The plotting style is as described in Figure \ref{fig:gas_strong}. Only in the most luminous nebular regions, we detected weak emission lines such as $\rm [Ne \, V]\lambda3427$, $\rm H\delta$, $\rm H\gamma$, $\rm [O \, III]\lambda4364$, and $\rm He \,II \lambda4687$.}
    \label{fig:gas_faint}
\end{figure*}

To measure the emission properties for each region, we jointly fit the strong and weak emission lines described above with Gaussian profiles using \texttt{LMFIT} \citep{2014zndo.....11813N}. For each region, all fitted lines share the same redshift and velocity width, but line fluxes are free parameters except for cases with line ratios set by atomic physics (e.g., $\rm [O \, III]\lambda4960$ and $\rm [O \, III]\lambda5008$). In most cases, a single set of Gaussians is enough to describe the emission line profiles, except for S3, S4, and B4 which require a second set of Gaussians to account for broader ($\sigma \approx 100{-}170 \, \rm km \, s^{-1}$) emission wings. Such emission wings are often seen around luminous quasars due to quasar-driven outflows \citep{1981ApJ...247..403H, 2013MNRAS.430.2327L, 2013MNRAS.436.2576L}, but the wings on S3, S4, and B4 may also be due to projection effects. We summarize the measurements for these regions, including their distances from the quasar, extraction radii, line fluxes, LOS velocities, and 1-D velocity dispersions, in Table \ref{tab:gas_pc}. We display strong and weak line spectra as well as their best-fit models in Figure \ref{fig:gas_strong} and Figure \ref{fig:gas_faint} respectively for a representative subset of the regions.

\section{Discussion}
\label{Dis}

\begin{figure}
    \centering
    \includegraphics[scale=0.5]{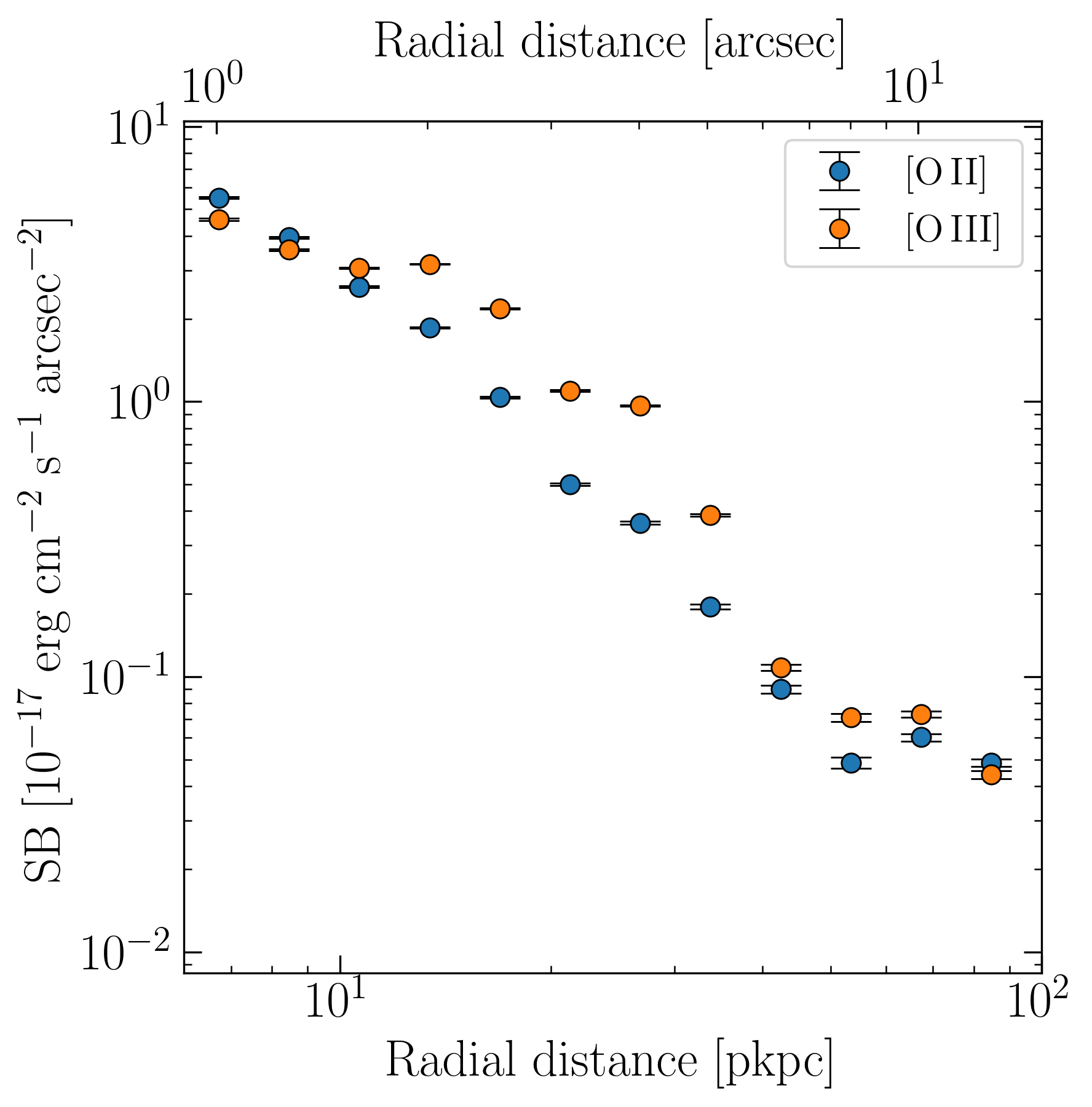}
    \caption{\textcolor{black}{Emission line surface brightness profile for the nebula around HE\,0238$-$1904. The [O\,II] and [O\,III] profiles are extracted over a velocity interval of $-600$ to $600 \, \rm km \, s^{-1}$, and are circularly averaged at different distances from the quasar centroid. The profile of [O\,II] declines smoothly as a function of radius, while the [O\,III] exhibits shallower drop due to the bright knots seen in the narrow-band images.}}
    \label{fig:RP}
\end{figure}

As discussed in Section \ref{GE}, the environment of HE\,0238$-$1904 is overdense and includes a massive galaxy group \textcolor{black}{or cluster}. Based on clustering studies, this environment is richer than those of most radio-quiet systems, but consistent with expectation for radio-loud ones. This demonstrates that radio-quiet systems like HE\,0238$-$1904 are diverse in terms of their host environment. Nevertheless, the lack of detected radio emission and amorphous morphology of the nebula suggests that it is not jet related. Considering that most published giant nebulae at $z < 1$ are in a rich environments, the presence of giant nebulae might be correlated with group properties. A larger sample size of quasars with wide IFS observations is required to investigate this possibility.

\textcolor{black}{Alternatively, such a rich environment can be explained by variable radio quasars. Quasars are capable of changing from radio-quiet to radio-loud or vice versa. \citet{2020ApJ...905...74N} found 26 sources showing radio variability over timescales of decades from the SDSS DR14 quasar catalog \citep{2018A&A...613A..51P} and the Wide-field Infrared Survey Explorer (WISE; \citealt{2010AJ....140.1868W}) R90 quasar catalog \citep{2018ApJS..234...23A}. These sources, once considered radio-quiet quasars, now meet the criteria for radio-loud ones. It implies that the probability that any particular radio-quiet quasar becomes radio-loud on the light-crossing timescale of the nebula is approximately $1\%$. However, the presence of a massive group and nebula mean that HE\,0238$-$1904 is not a representative quasar and so may be more likely to transition to radio-loud relatively soon. On the other hand, the possibility that HE\,0238$-$1904 was previously radio-loud and is now radio-quiet is harder to address since such transitions are not well studied.}

\textcolor{black}{In the following subsections, we discuss insights into the physical origins and state of the giant nebula which includes analyses of density and ionization-state sensitive diagnostic emission lines. Several of these analyses require priors on the dust content and density of the gas. To investigate dust content, we estimate Balmer line ratios, and find $\rm H\delta/H\gamma$ ratios of $\approx 0.55$. These ratios are consistent with Case B recombination \citep{2006agna.book.....O} in the absence of dust. To obtain density estimates, we infer emission measure of the nebula from the surface brightness of $\rm H \beta$ following \cite{2019ApJ...878L..33C}. Assuming $\rm H\alpha/H\beta\approx3$, a clumping factor of 1, and length-scale $30 \rm \, pkpc$, we found an electron density of $\log(n_{\rm e}/{\rm cm}^{-3})\approx -1$. However, this density estimate has a large uncertainty and is effectively a lower limit due to the assumption of a unity clumping factor. }

\subsection{Origin of the Nebular Gas}
\label{subsection:origin}
Giant nebulae can be produced via ram pressure and tidal stripping, AGN and stellar feedback, or filamentary accretion. The nebula around HE\,0238$-$1904 is unlikely to arise from a jet-driven outflow given the fact that the quasar is radio-quiet and exhibits no detectable radio jet. While S3 and S4 exhibit broad emission wings, most regions are well characterized by a single Gaussian profile with narrow velocity dispersion ($\sigma < 120 \, \rm km \, s^{-1}$; see Table \ref{tab:gas_pc}). \textcolor{black}{These} quiescent kinematics \textcolor{black}{are} inconsistent with the broad velocity dispersion expected from radio-quiet AGN and stellar feedback \citep{2013MNRAS.436.2576L, 2019Natur.574..643R}. In addition, the morphology is inconsistent with expectations for filamentary accretion \citep{2022ApJ...940L..40J}. On the other hand, the nebula is spatially and kinematically coincident with likely interacting galaxies in the field of HE\,0238$-$1904, suggesting that stripping from interactions is likely responsible for most of the nebula with possible subdominant contributions from outflows.

The nebula spatially surrounds the Host, G1, G3, G4, and G5, and extends to the South West of the \textcolor{black}{quasar} to a projected distance of $d \sim \rm 70 \, pkpc$. This spatial coincidence suggests that the nebula likely arises from interaction-related stripping. The dwarf galaxies G3 and G5 show a \textcolor{black}{possible} tidal-tail-like structure as shown in panels (e) and (h) of Figure \ref{fig:gas}, suggesting that this part of the nebula might be created from tidal stripping. In addition to this, the emission maps on larger scales resemble a head-tail morphology with the head around the quasar and with the tail extending to the South West of the quasar. Head-tail morphologies are commonly seen in nebulae originated from ram pressure stripped ISM \citep[e.g.,][]{2016AJ....151...78P, 2019A&A...631A.114B, 2019ApJ...878L..33C}. \textcolor{black}{Interestingly, while the nebula exhibits a head-tail morphology, it does not exhibit multiple filaments like some ``jellyfish'' galaxies observed in the optical line emission. Instead, it resembles the smoother emission profile sometimes seen in ram-pressure debris observed in H\,I 21-cm \citep[][]{2017MNRAS.464..957H}.} There are two plausible explanations for ram pressure stripping in the environment of HE\,0238$-$1904. First, the nebula may arise from stripping of the quasar host's ISM and CGM if it is falling into the richer, redshifted group and passing through the associated hot halo. Second, dwarf galaxies may have travelled through the hot halo of the \textcolor{black}{massive group} from West to East, leaving their ram pressure stripped ISM and CGM behind along their path. 

The discovery of a giant nebula requires both the presence of gas and its positioning within quasar's ionization cone. However, due to projection effects, the relative position between the quasar and the nebula remains uncertain. The two previously mentioned hypotheses provide potential frameworks. (1) If the gas results from stripping of the quasar host's ISM, the nebula is likely to surround the quasar. In this case, it will naturally be illuminated by the quasar. Alternatively (2) if the nebula arises from the stripped CGM/ISM of other galaxies in the overdensity, the gas will be widely distributed throughout the groups and more distant from the quasar. Only a fraction of this gas might coincidentally fall within the quasar's ionization cone, consistent with the large opening angle suggested by \citet{2013ApJ...775L...3T, 2016ApJ...830..120B, 2018ApJ...861..122S, 2020MNRAS.495.1874D}.

To examine between these scenarios, we show the surface brightness profiles of [O\,II] and [O\,III] made with \texttt{Photutils} \citep{larry_bradley_2023_7946442} in Figure \ref{fig:RP}. The profile of [O\,II] declines smoothly as a function of radius, and plateaus at $\approx \rm 50 \, pkpc$. In contrast, the [O\,III] profile exhibits shallower drop due to the bright knots seen in the narrow-band images. The plateau in the [O\,II] profile corresponds to the head-tail morphology of the nebula, and the bright knots hints at a dwarf-related origin for part of the nebula. Collectively, the [O\,II] and [O\,III] profiles suggest a complex scenario. The centroids of narrow-band [O\,II] and [O\,III] surface brightness maps are 10 and 19 pkpc away from the quasar respectively, an alignment to within $15\%$ of the size of the nebula. This coincidence could be explained if the gas surrounds the quasar or if the quasars ionization cone is fairly well centered on our LOS. However, the significant contributions of individual dwarf galaxies to the [O\,III] surface brightness profile underscore the challenge in precisely determining the nebula's position relative to the quasar. Consequently, it is plausible that both scenarios (1) and (2) contribute to the nebula.

The giant nebulae around HE\,0238$-$1904 was independently discovered and reported by \cite{2023ApJ...943L..25Z}. They attributed the gas to a superbubble driven by the quasar based on an apparent large velocity shift between the nebula and the quasar redshift and as well as broad line widths reported near the quasar.
However, the large velocity shift is due to the reliance on an older, Mg\,II-based redshift of $z=0.631$, which is $\approx +500\ {\rm km\,s^{-1}}$ from our [O\,II]-based redshift of $z=0.6282$.
\textcolor{black}{Rather than relying on a redshift estimate from the literature, we measured the quasar redshift and kinematics of the giant nebula from the same MUSE dataset to avoid any systematic uncertainty due to wavelength calibration errors. Moreover, quasar redshifts based on [O\,II] are generally more accurate than those measured from Mg\,II due to the narrowness of the line and lack of blueshifted wings on [O\,II]. In particular, quasar redshifts measured from [O\,II] trace the underlying quasar host redshifts measured in stellar absorption to within $\approx \pm 20\ {\rm km\,s^{-1}}$ \citep[][]{2010MNRAS.405.2302H}. Finally, our redshift estimate of $z=0.6282$ is more consistent with the centroid of the broad H$\beta$ line, aligns with the peak of the quasar's [O\,III] emission line, and matches a more recent Mg\,II-based redshift of $z=0.628$ from the UV-bright Quasar Survey \citep[][]{2016AJ....152...25M}. Furthermore, we measured significantly narrower line widths near the quasar. This is likely due to our removal of [O\,III] and [O\,II] emission from the unresolved narrow-line emission region of the quasar while \cite{2023ApJ...943L..25Z} only removed emission from the broad-line region. In summary, the modest velocity shifts and largely narrow emission line widths are consistent with much of the gas originating from interactions with more minor possible contributions from an outflow. When using the updated quasar redshift and quasar-light subtracted datacube, we find no evidence for a fast, quasar driven superbubble in the system.}

\begin{table*}
	\centering 
	\caption{Summary of nebula regions in the Field of HE\,0238$-$1904.}
	\label{tab:gas_pc2}
	\begin{threeparttable}
    \setlength{\tabcolsep}{10pt} 
    \renewcommand{\arraystretch}{1.2} 
	\begin{tabular}{lccc}
		\hline
		ID & $\log (n_{\rm e, [O \, II]} /\mathrm{cm}^{-3})$\tnote{a} & $\log (n_{\rm H, Cloudy} /\mathrm{cm}^{-3})$\tnote{b} & ${\rm log}(U_{\rm Cloudy})$\tnote{c} \\
		& & & \\
		\hline
		S1 & $<1.6 $ & $1.6^{\hspace{0.2mm}+\hspace{0.1mm}0.1}_{-0.1}$ & $-2.2^{-0.1}_{\hspace{0.2mm}+\hspace{0.15mm}0.1}$ \\
		S2 & $<1.7$ & $1.7^{\hspace{0.2mm}+\hspace{0.1mm}0.1}_{-0.1}$  & $-2.1^{-0.1}_{\hspace{0.2mm}+\hspace{0.15mm}0.1}$ \\
		S3 & ... & ... & ... \\
		S4 & ... & ... & ... \\
		S5 & $<1.6$ & $4.2^{\hspace{0.2mm}+\hspace{0.1mm}0.2}_{-0.3}$ & $-3.0^{-0.2}_{\hspace{0.2mm}+\hspace{0.15mm}0.3}$ \\
        S6 & $\hspace{7.3mm} 1.8^{\hspace{0.2mm}+\hspace{0.1mm}0.1}_{-0.1}$ & $2.7^{\hspace{0.2mm}+\hspace{0.15mm}0.1}_{-0.1}$ & $-2.5^{-0.1}_{\hspace{0.2mm}+\hspace{0.15mm}0.1}$ \\
        S7 & $<1.9$ & $3.0^{\hspace{0.2mm}+\hspace{0.1mm}0.3}_{-0.3}$ & $-3.2^{-0.3}_{\hspace{0.2mm}+\hspace{0.15mm}0.3}$ \\
        S8 & $<1.3$ & $3.5^{\hspace{0.2mm}+\hspace{0.1mm}0.1}_{-0.2}$ & $-3.3^{-0.1}_{\hspace{0.2mm}+\hspace{0.15mm}0.2}$ \\
        S9 & $<2.3$ & $4.1^{\hspace{0.2mm}+\hspace{0.1mm}0.2}_{-0.3}$ & $-3.5^{-0.2}_{\hspace{0.2mm}+\hspace{0.15mm}0.3}$ \\
        S10 & $<1.4$ & $3.6^{\hspace{0.2mm}+\hspace{0.1mm}0.2}_{-0.2}$ & $-3.3^{-0.2}_{\hspace{0.2mm}+\hspace{0.15mm}0.2}$ \\
        B1 & $<2.8$ & $2.1^{\hspace{0.2mm}+\hspace{0.1mm}0.1}_{-0.2}$ & $-2.7^{-0.1}_{\hspace{0.2mm}+\hspace{0.15mm}0.2}$ \\
        B2 & $<1.2$ & $2.9^{\hspace{0.2mm}+\hspace{0.1mm}0.1}_{-0.3}$ & $-3.4^{-0.1}_{\hspace{0.2mm}+\hspace{0.15mm}0.3}$ \\
        B3 & $<2.5$ & $1.9^{\hspace{0.2mm}+\hspace{0.1mm}0.1}_{-0.2}$ & $-2.8^{-0.1}_{\hspace{0.2mm}+\hspace{0.15mm}0.2}$ \\
        B4 & ... & ... & ... \\
		\hline \\
	\end{tabular}
	\begin{tablenotes}
	    \footnotesize
        \item \textbf{Notes.}
        \item[a] Number density measurement or $95$\% upper limit measured from $\rm [O \,II]\lambda3729/[O \,II]\lambda3727$.
        \item[b] Number density inferred from \texttt{Cloudy} simulation described in the text.
        \item[c] Best-fit ionization parameter computed by \texttt{Cloudy} simulation.
    \end{tablenotes}
	\end{threeparttable}
\end{table*}

\subsection{Physical Conditions of the Emitting Gas}
Previous studies of giant nebulae have attributed the ionization of the gas to ionizing photons from AGN, shocks, and young stellar populations \citep[e.g.,][]{2018ApJ...869L...1J, 2019Natur.574..643R, 2019ApJ...878L..33C, 2021MNRAS.505.5497H, 2023Sci...380..494Z}. The presence of the quasar suggests the source of ionization is AGN-related. To study the physical conditions of the gas, we measured the the density- and temperature-sensitive $\rm [O \, II]\lambda3729/[O \, II]\lambda3727$ and $\rm [O \, III]\lambda4364/[O \, III]\lambda5008$ line ratios as well as ionization state-sensitive strong and weak line ratios in each region. These line ratio measurements are reported in Table \ref{tab:gas_pc} and \textcolor{black}{a [O\,III]/[O\,II] map} is shown in panel (c) of Figure \ref{fig:gas}. We discuss these measurements and their implications in the following three subsections.

\subsubsection{Direct Density and Temperature Estimates}
With spectral coverage of $\rm [O \, II]\lambda3727$, $\rm [O \, II]\lambda3729$, $\rm [O \, III]\lambda4364$, and $\rm [O \, III]\lambda5008$, we can directly measure electron density ($n_{\rm e}$) and temperature ($T_{\rm e}$), as discussed in \cite{2006agna.book.....O}. The $\rm [O \, II]$ doublet is a good density estimator because the difference in excitation energy between these two upper states is small so that the relative population in the two states is determined by electron density and is insensitive to temperature. In contrast, the $\rm [O \, III]$ doublet upper states have a larger excitation energy difference, making the populations of these states mainly sensitive to electron temperature and insensitive to electron density. Electron number densities from the $\rm [O \, II]$ doublet are reasonable proxies for the overall densities of ionized nebulae because $\rm H$ and $\rm O$ share similar ionization energies of $13.6 \, \rm eV$.

To translate line ratios into physical conditions, we used  \texttt{Pyneb} \citep{2015A&A...573A..42L} which predicts the [O\,II] and [O\,III] line ratios  at a given density and temperature by solving the detailed balance equation for an $n$-level atom. We fit the measured line ratios with \texttt{Pyneb} models by performing Markov chain Monte Carlo (MCMC) analysis with \texttt{emcee} \citep{2013PASP..125..306F}, and inferred physical conditions from the resulting posteriors. We report the densities in Table \ref{tab:gas_pc2}, though we omit measurements in cases where the S/N or broad line width results in poorly constrained conditions.

For all regions where the $\rm [O \, II]$ doublet is resolved, the line ratio is in the low density limit except S6. We therefore report 95$\%$ upper limits in density for all but S6. The inferred electron number density upper limits range from $1.2 < \log(n_{\rm e, [O \, II]} / \mathrm{cm^{-3}}) < 2.8$, with a median of $\log(n_{\rm e, [O \, II]} / \mathrm{cm^{-3}}) < 1.6$. These density upper limits are consistent with gas arising from ionized ISM \citep{2011piim.book.....D} or CGM.  We detected $\rm [O \, III]\lambda4364$ in only three luminous regions, S1, S2, and S6. The inferred temperatures for S1, S2, and S6 are $\log(T/ \mathrm{K})\approx 4.2$, $4.2$, and $4.1$ respectively.

\subsubsection{Indirect Density Estimates from Photoionization Simulations}
\label{IDEFPS}
Under the assumption that the nebula is ionized by the quasar, its ionization states are set by the luminosity of the quasar, density of the gas, and distance from the quasar, with secondary effects from metallicity and ionizing spectral shape. With an estimate of the quasar's luminosity and assuming projection effects are negligible, the density structure of the gas can be inferred from measured line ratios \citep[see][]{2019MNRAS.483.5188C}. Studies of high redshift quasar nebulae found ionization states can only be explained by a density of $\log(n_{\rm H} / \mathrm{cm^{-3}}) \approx 1.9$, significantly higher than expected CGM/IGM densities, or alternatively by a broad density distribution \citep[see][]{2019MNRAS.483.5188C}. At low redshift, this kind of scenario can be further explored with insight from rest-optical lines to compare ionization-based densities with more direct density estimates from the $\rm [O \, II]$ doublet.

To infer the physical conditions from the line ratios in Table \ref{tab:gas_pc}, we ran photoionization simulations for each region with \texttt{Cloudy} version C17.03 \citep{2017RMxAA..53..385F}. We modelled the quasar's radiation field using a power law ($I \propto \nu^{\alpha}$) between 0.37 and 73.5 $\rm Ryd$, with $\alpha$ between $-1.8 < \alpha < 0$ following \citet{2004ApJS..153....9G} but extending to a higher $\alpha$. We set the modeled quasar luminosity at $1\, \rm Ryd$ using direct measurement of the monochromatic UV luminosity from COS. For the gas, we adopted single density and single metallicity models, with 
density of $-2< \log(n_{\rm H} / \mathrm{cm}^{-3})< 4.6$ and metallicity of $-1.5 < \log(Z/Z_{\odot})< 0.5$. \textcolor{black}{We chose this metallicity range to cover the characteristic metallicities of the cool CGM around massive elliptical galaxies \citep{2019MNRAS.484.2257Z} but extended it to higher metallicity in case some gas has ISM origins. } Due to limited ion coverage, metallicity and $\alpha$ are degenerate in some cases, so we treated them as nuisance parameters and focused on inferred densities. We note that there is relatively little degeneracy between density and metallicity except at high metallicities of $ \log(Z/Z_{\odot})>0.2$ when increased cooling from metal lines begins to substantially change the equilibrium temperature.

For each region, we \textcolor{black}{conducted} these models in grids with a step of $0.2$ \textcolor{black}{dex in density and metallicity, and 0.2 in $\alpha$.} We then interpolated these models with the \texttt{RegularGridInterpolator} function from \texttt{scipy.interpolate} \citep{2020SciPy-NMeth} within these ranges after checking for convergence. Finally, we ran \texttt{emcee} to estimate posteriors given the measured line ratios and uncertainties. We verified the quality of the fits by comparing the posteriors of the model line ratios with the measured line ratios using violin plots shown in Figure \ref{fig:violin}. The violin plots verify that the ionization-state-sensitive line ratios (shown in the middle panels) are consistent with the measured line ratios. \textcolor{black}{The best-fit $\alpha$ values for most regions are within $-1.0 < \alpha < -0.6$, somewhat greater than ones given in \citet{2004ApJS..153....9G}. Inferred metallicities for S1, S2, and S6, with He II and [Ne V] detections, are well-constrained to be $-0.2<\log(Z/Z_{\odot})<0.2$}. The \textcolor{black}{densities inferred} from these photoionization simulations range from $\log(n_{\rm H, Cloudy} / \mathrm{cm}^{-3}) = 1.6$ to $4.2$ and are reported in the right column of Table \ref{tab:gas_pc2}, though we stress that these densities neglect \textcolor{black}{potential} quasar variability and projection effects.

\subsubsection{Comparison of the Density Estimates}
\begin{figure*}
    \centering
    \includegraphics[scale=0.5]{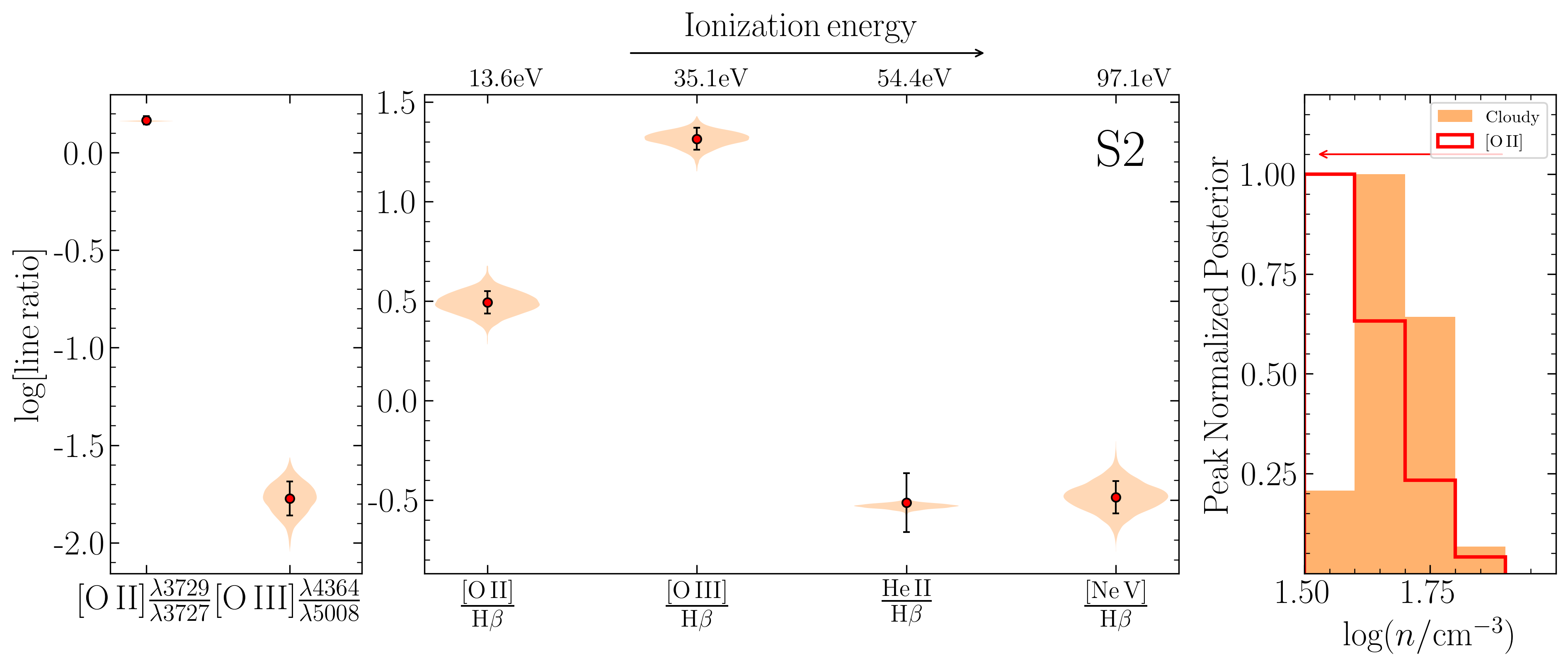}
    \includegraphics[scale=0.5]{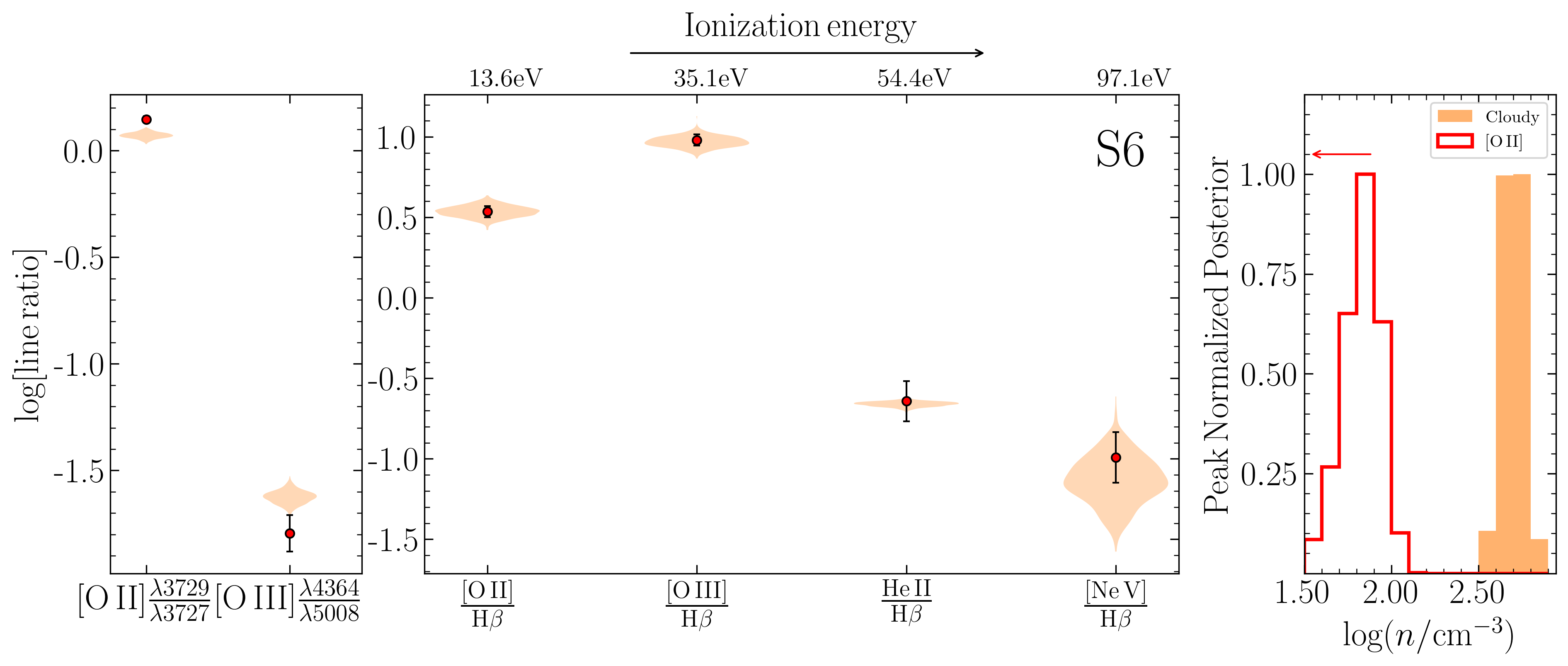}
    \includegraphics[scale=0.5]{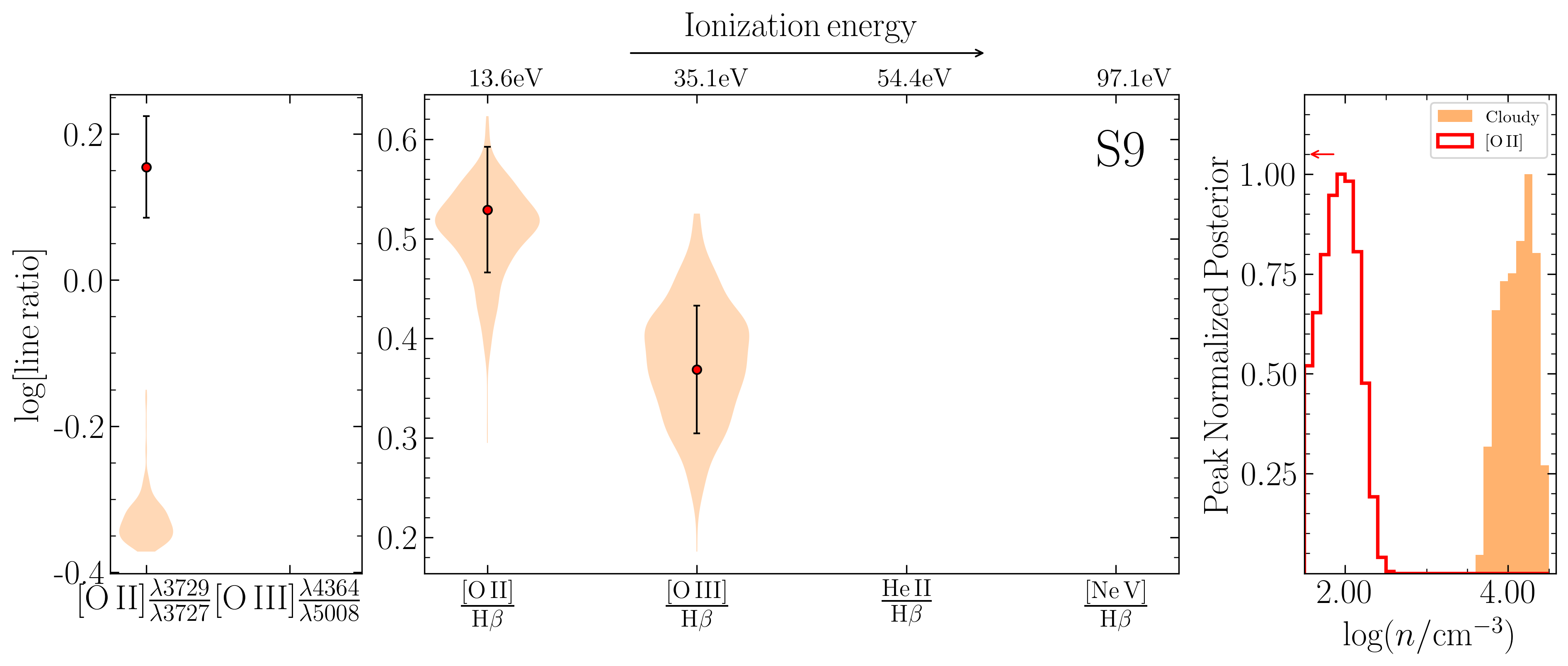}
    \caption{Examples of violin plots and density posteriors for nebular regions. For each subfigure, the left panel shows the density- and temperature-sensitive line ratios from the flux measurements as red points with error bars and the photoionization model posteriors are shown in orange. The middle panel shows the ionization-sensitive line ratios from the flux measurements as red points with error bars and the photoionization model posteriors in orange. The right panel shows the density posterior from both direct (red histogram) and indirect density estimates (orange filled histogram). The density posterior inferred from the [O\,II] doublet extends below the plotted range as indicated by the red arrows. For S6 and S9, photoionization simulations are able to reproduce the ionization-sensitive line ratios, but not the density-sensitive line ratio, indicating that the ionization-state-based density estimates are inconsistent with the direct [O\,II]-based density measurements.}
    \label{fig:violin}
\end{figure*}

Previous photoionization-based estimates of the density of quasar nebulae at high-redshift found unexpectedly high densities, close to or \textcolor{black}{exceeding typical densities for the ISM}, despite being \textcolor{black}{measured} on CGM/IGM scale \citep{2019MNRAS.483.5188C}. The ionization sensitive line ratios of the nebula around HE\,0238$-$1904 also imply high photoionization-based densities of $1.6< \log(n_{\rm H, \, Cloudy} / \mathrm{cm}^{-3})< 4.2$. However, the more direct $\rm [O \, II]$-based densities are inconsistent with and significantly smaller \textcolor{black}{than} the photoionization-based densities for most regions as shown in Table \ref{tab:gas_pc2}. To better demonstrate this inconsistency, Figure \ref{fig:violin} shows both the measured line ratios and the posteriors inferred from the photoionization models for S2, S6, and S9. The ionization-state-sensitive line ratios are consistent with the model posteriors for all three regions, while the $\rm [O \, II]$ line ratios are highly discrepant for S6 and S9. The right panel of each subfigure shows the density posteriors from both direct and indirect density estimates.

As shown in Table \ref{tab:gas_pc2}, we found that all regions with photoionization-based density estimates except S1, S2, B1, and B3 have a large ($1{-}2$ dex) discrepancy when compared to the [O\,II] doublet-based densities. In the most extreme case, S5, the two density estimates are off by $2.6$ dex or a factor of $400$. In principle, the inferred density mismatch could be explained by a non-uniform density distribution if the [O\,II] arises from less dense gas than the other emission lines. To test whether a more complicated density structure could explain the density mis-match, we modeled the emitting gas as a multi-phase system consisting of one low density component and one high density component with the relative contribution of each treated as an additional free parameter. This model successfully reproduces the observed emission-line ratios, and the density inferred for the high density component matches the single-phase model results. Furthermore, the posteriors of the two-component model indicate that the high density component dominates the $\rm [O \, II]$ emission. Therefore, a two-phase model cannot explain the density discrepancy between the direct [O\,II]-based density measurements and the ionization-state-based density estimates.

\textcolor{black}{To test if a broad, continuous density distribution can explain the discrepancy, we modelled the emitting gas with a log-normal density distribution \citep[see][]{2019MNRAS.483.5188C}. A log-normal distribution is defined as
\begin{equation}
    {\rm PDF}(n)\text{d}n = \frac{1}{\sqrt{2 \pi}\sigma} {\rm exp} \Big[- \frac{[\text{ln}(n) - \text{ln}(\mu)]^2}{2 \sigma^2} \Big]\text{dln}(n)
\end{equation}
where $\sigma$ is the dispersion and $\mu$ is the mean density. We started with calculating emission line emissivity in an extended \texttt{Cloudy} model grid, similar to ones discussed in Section \ref{IDEFPS}. We then computed the predicted line ratios for a log-normal density distribution by interpolating \texttt{Cloudy} models and integrating over the PDF. Our results show that a log-normal distribution with a large $\sigma$ can reproduce the ionization-sensitive line ratios, but the log-normal models predict that the [O\,II] emission arises from dense gas, resulting in [O\,II] line ratios of $\log(\frac{\lambda3729}{\lambda3727}) = {-}0.4 \, {\rm to} \, {-}0.1$, inconsistent with the observed ratios of $\log(\frac{\lambda3729}{\lambda3727}) > 0.1$. Therefore, a broad density distribution is unlikely to reconcile the density discrepancy.}

Alternatively, projection effects can also result in disagreement between the two density estimates. However, assuming that the gas is randomly and approximately spherically distributed around the quasar, the projected distance is unlikely to be much smaller than the radial distance between the quasar and the nebula. For example, producing a factor of $400$ mismatch in density requires the radial distance to be 20 times larger than the projected distance. While such projection effects are possible in principle, the required contrived geometry is unlikely.

In principle, the discrepancy in density could be explained if the nebula is not directly ionized by the quasar due to obscuring dust or translucent clouds blocking its light from reaching this gas. Filtering the quasar's radiation through dust would soften the incident ionizing radiation field. However, the best-fit $\alpha$ values from our photoionization analysis suggests a hard ionizing spectrum for almost all regions. The hard inferred ionizing spectrum is inconsistent with expectations from a quasar SED filtered through dust clouds. 

Alternatively, translucent clouds of moderate optical thickness to ionizing photons can also filter the quasar's radiation. Depending on the density and the physical size, these clouds could produce distinct line ratios as a function of depth into the cloud \citep{2013MNRAS.430.2327L}. Typically, the outer parts of the cloud produce no significant [O\,II] or [O\,III] emission because oxygen is highly ionized. However, $\rm H\beta$ is a recombination line and so a non-negligible fraction of the $\rm H\beta$ emission arises from outer parts of the cloud that do not emit in [O\,II] or [O\,III]. As a result, translucent regions are expected to have stronger $\rm H\beta$ emission than $\rm [O \, II]$ and $\rm [O \, III]$. Yet, none of the nebular regions have such $\rm [O \, III]/H\beta$ ratio. If these translucent clouds exist around HE\,0238$-$1904, they therefore must be blended with optically thick clouds due to seeing conditions and projection effects. The presence of unresolved translucent clouds could be investigated by observing the nebula with higher spatial resolution instruments such as NIRSpec on the JWST or with adaptive optics from the ground. Nevertheless, while translucent clouds may help reconcile the density discrepancy in some cases, moderate optical depth clouds can only absorb a modest portion of the quasar's radiation. Therefore, it is unlikely to explain the largest density discrepancies.

\begin{figure*}
    \centering
    \includegraphics[scale=0.35]{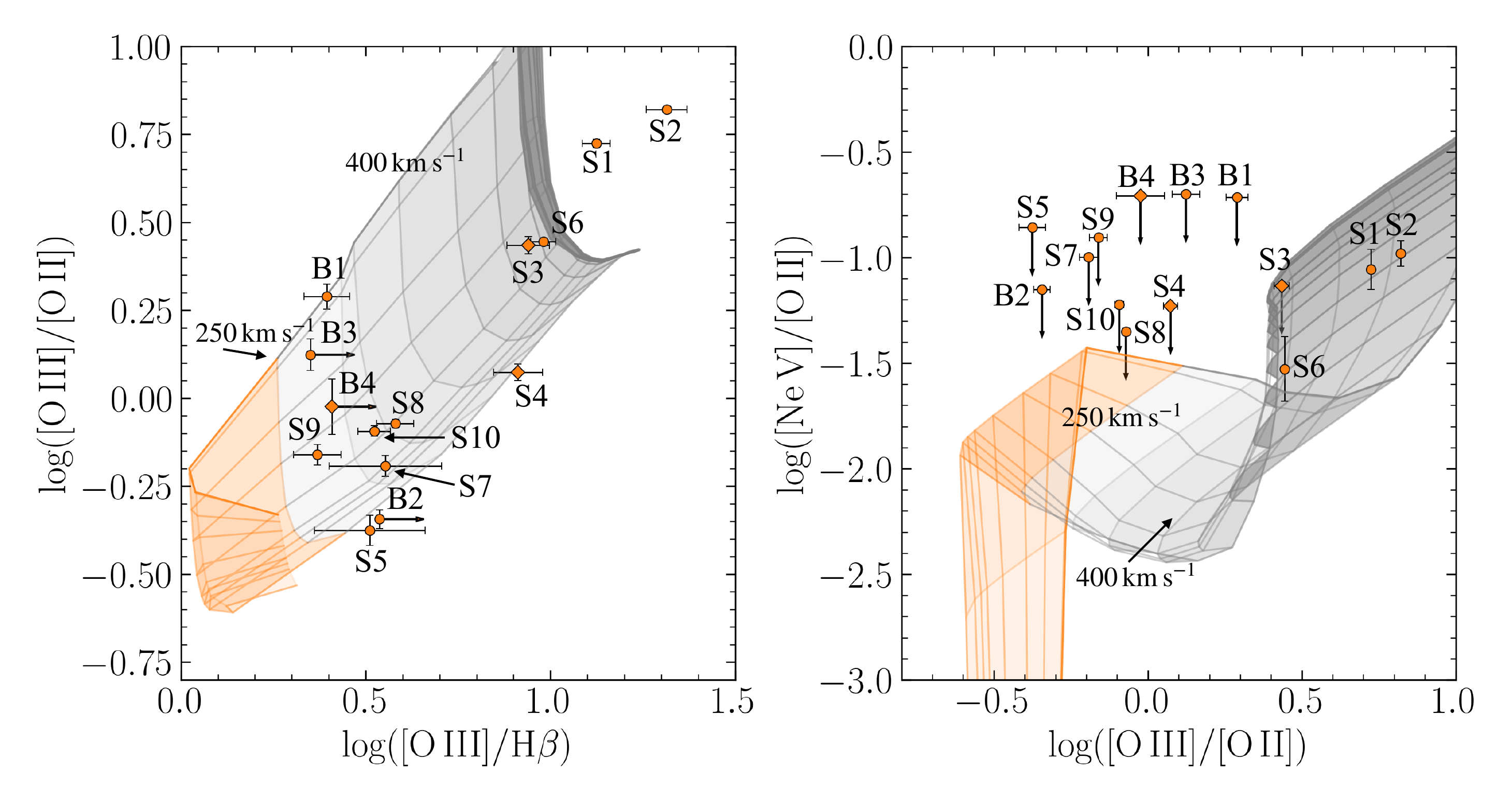}
    \caption{The emission line diagnostic diagrams log([O III]$\lambda 5008$/[O II]$\lambda\lambda 3727, 3729$) versus log([O III]$\lambda5008$/H$\beta$) and log([Ne V]$\lambda3427$/[O II]$\lambda\lambda 3727, 3729$) versus log([O III]$\lambda 5008$/[O II]$\lambda\lambda 3727, 3729$) for nebular regions. Line ratios are shown as orange points with error bars, or shown as $3\sigma$ upper limits for non-detections. For S3, S4, and B4, total line ratios (main+wing) are shown as orange diamonds with error bars, or with upper limits. We note that S3, S4, and B4 might have large uncertainty due to multiple components detected within $150 \rm \, km \, s^{-1}$. We compare these line ratios with the fast radiative shock models (shock plus precursor) from \citet{2008ApJS..178...20A}. Emission-line ratio grids with solar metallicity, a preshock density of $n_{\rm e} = 100 \, \rm cm^{-3}$, and a magnetic field strength of $B= 0.001{-}100 \rm \, \mu G$ are shown in orange and grey for a shock velocity of $100{-}250 \rm \, km \, s^{-1}$ and $250{-}1000 \rm  \, km \, s^{-1}$ respectively. Given the quiescent kinematics of the nebular regions, shocks are unlikely to be the source of ionization in most cases.}
    \label{fig:shck}
\end{figure*}

On the other hand, the ionization of the nebulae could be due to young stellar populations \citep{2015RMxAA..51..103M} or fast shocks \citep{2008ApJS..178...20A}. However, there is no evidence of extended star-formation in rest-frame $u$-band images of the system formed from the MUSE datacube. To investigate the possibility of fast shocks, we show two emission line diagnostic diagrams overlaid with shock models in a grid of shock velocity and magnetic field strength in Figure \ref{fig:shck}. Producing the observed [O\,III]/[O\,II] and [Ne\,V]/[O\,II]\footnote{We note that [Ne\,V]/[Ne\,III] as a better shock tracer cannot be used due to [Ne\,III]$\lambda$3869 is severely contaminated by skylines.} ratios requires shock velocities of  $v_{\rm shock}> 250 \, \rm km \, s^{-1}$ \citep{2008ApJS..178...20A}. These shock velocities are greater than the LOS velocity and velocity dispersion of the nebula in nearly all locations, even after accounting for projection effects.
For example, some regions (S1 and S2) would require shock velocities exceeding $1000 \, \rm km \, s^{-1}$ and most regions (S3, S4, S6, S8, S10, B1, B2, B3, and B4) would require $>300\!-\!400 \, \rm km \, s^{-1}$, making them unlikely to be ionized by shocks. On the other hand, while the observed line ratios of S5, S7, and S9 favor AGN photoionization,  large uncertainties in their $\rm H\beta$ flux can accommodate shocks with velocities as low as $200 \, \rm km \, s^{-1}$. This would alleviate the density discrepancy in these three regions. However, for most regions, the shock velocity required to reproduce the observed line ratios exceeds velocities observed in the system. Shocks are therefore unlikely to explain the density discrepancy in most cases.

Perhaps more likely, the difference in the density estimates could be due to quasar variability \citep{1977ApJ...213....8R}. Quasar variability is directly observed on timescales of decades \citep{2022MNRAS.514..164S}. \textcolor{black}{Observations of ``changing-look'' AGN, light echoes, and quasar proximity zones suggest the average episodic lifetime of quasars may range from $10^4$ to $10^7$ years and AGN episodes may be highly clustered} \textcolor{black}{\citep[e.g.,][]{2004ApJ...610..105S, 2008ApJ...676..816G, 2008MNRAS.391.1457K, 2013ApJ...775L...3T, 2014ApJ...784...42S, 2015MNRAS.451.2517S, 2017ApJ...849..102C, 2018ApJ...861..122S, 2021ApJ...921...70S}}. Therefore, each region of the nebula around HE\,0238$-$1904 may experience a drastically different radiation field from the quasar, depending on the light travel time. For example, S5 and S6 are at a projected distance of $10$ to $\rm 20 \, pkpc$ from the quasar, respectively, and their line ratios can be explained if the quasar was $400$ and $10$ times less luminous than currently observed. In contrast, S1 and S2 are at a projected distance of $\approx \rm 40 \, pkpc$ from the quasar, and their properties can be explained if they received ionizing radiation consistent with the current luminosity of the quasar. We confirmed that quasar variability could explain the ionization state and $\rm [O \, II]$ ratio by re-running \texttt{Cloudy} models and MCMC analysis after significantly decreasing the quasar luminosity.

\section{Summary and Conclusions}
\label{SC}
In this paper, \textcolor{black}{we presented the first comprehensive analysis of a giant nebula around a radio-quiet quasar at $z<1$ based on MUSE observations of the field of HE\,0238$-$1904}. The wide FoV, high spatial sampling, and wide wavelength coverage enabled us to investigate the origin and the physical condition of the group and gaseous environment with a spatially resolved analysis of the morphologies, kinematics, and nebular photoionization properties. Our finding can be summarized as follows.

\begin{enumerate}
    \item We found that HE\,0238$-$1904 resides in an overdense environment containing two potentially merging galaxy groups based on spatial distribution and kinematics. This includes a less rich, blueshifted group with $12$ galaxies and a richer, redshifted group with $22$ galaxies. Assuming the more massive group is virialized, its dynamical mass is $M_{\rm dyn} \sim 4 \times 10^{13}{-}10^{14}\ {\rm M_{\odot}}$. Such a massive, rich environment is unusual for a radio-quiet quasar, which \textcolor{black}{typically resides in a halo with a mass of $\sim 3 \times 10^{12}\ {\rm M_{\odot}}$ \citep{2009ApJ...697.1656S}.}
    
    \item We identified a giant nebula covering a projected area of $\approx 5000\ {\rm kpc}^2$ around HE\,0238$-$1904 emitting strongly in $\rm [O \, II]$, $\mathrm{H}\beta$, and $\rm [O \, III]$. The nebula has an irregular morphology with a spatial trend in kinematics where the region North of the quasar is redshifted and the region South of the quasar is mainly blueshifted relative to the quasar. The southern region is spatially coincident with four dwarf galaxies.
    
    \item The coincidence with nearby galaxies suggests that it arises from stripping of ISM or CGM, which is consistent with its morphology and largely narrow LOS velocity dispersion. In addition, the nebula shows a head-tail morphology with the head near the quasar and with the tail extending toward South West of the quasar. The head-tail structure may originate from ram pressure if the quasar and the surrounding nebula are infalling toward the massive galaxy group to the North East. However, we note there are some small regions at $d \approx 20 \, \rm pkpc$ from the quasar that have broader emission wings, perhaps suggesting an outflow origin.
    
    \item To better characterize the physical conditions of the nebula, we measured \textcolor{black}{the fluxes of} strong and weak emission line fluxes. The inferred electron number density upper limits from the $\rm [O \, II]$ doublet range from $\log(n_{\rm e, [O \, II]} / \mathrm{cm^{-3}}) < 1.2$ to $2.8$, with a median of $\log(n_{\rm e, [O \, II]} / \mathrm{cm^{-3}}) < 1.6$. These density upper limits are consistent with ISM or CGM origin. However, densities inferred from photoionization models are often inconsistent with the $\rm [O \, II]$-based density upper limits, reaching values of up to $400$ times higher.
    
    \item The disagreement in density estimates is \textcolor{black}{unlikely to be due to density inhomogeneities, but can be explained by} quasar variability, if the quasar varied significantly on timescales of $10^4$ to $10^5$ years. This finding suggest that long-term quasar variability should be included when considering ionization-based inferences into the physical conditions of giant nebulae around quasars.
\end{enumerate}
\textcolor{black}{The possibility of significant quasar variability on timescales of $10^4$ to $10^5$ years has implications far beyond accretion disk physics in the central engine. In particular, significant fluctuations on these timescales can result in out-of-equilibrium conditions in the low density circumgalactic medium due to the long recombination time of low density gas \citep[][]{2013MNRAS.434.1063O, 2017MNRAS.471.1026S}. Indeed, such AGN ``flickering'' may be responsible for strong O\,VI absorption observed around Milky Way-like galaxies at low redshift \citep[][]{2018MNRAS.474.4740O}.} The recent and upcoming commissioning of new IFSs on large telescopes, such as LLAMAS \citep{2020SPIE11447E..0AF}, IFUM \citep{2022SPIE12184E..5PM}, Blue MUSE \citep{2019vltt.confE..24R}, and MIRMOS \citep{2020SPIE11447E..1EK}, will continue to drive further discoveries of giant nebulae which could be followed up with IFS like HARMONI \citep{2022SPIE12184E..20T} on future, 30-meter class telescopes, \textcolor{black}{extending similar insights to higher redshifts and fainter systems.}

\section*{Acknowledgements}
SDJ and ZQL acknowledge partial support from HST-GO- 15280.009-A, HST-GO-15298.007-A, HST-GO-15655.018-A, and HST-GO-15935.021-A. JIL is supported by the Eric and Wendy Schmidt AI in Science Postdoctoral Fellowship, a Schmidt Futures program. SC gratefully acknowledges support from the European Research Council (ERC) under the European Union’s Horizon 2020 Research and Innovation programme grant agreement No 864361. This paper is based on observations from the European Organization for Astronomical Research in the Southern Hemisphere under ESO (PI: J. Schaye, PID: 094.A-0131(B) \& 096.A-0222(A)), and the NASA/ESA Hubble Space Telescope (PI: L. Straka, PID: 14660; PI: J. Green, 11541; PI: S. Penton, PID: 12505). Additionally, this paper made use of the NASA/IPAC Extragalactic Database, the NASA Astrophysics Data System, \texttt{Astropy} \citep[][]{2022ApJ...935..167A}, \texttt{Aplpy} \citep{aplpy2012}, and \texttt{Photutils} \citep{larry_bradley_2023_7946442}.

\section*{DATA AVAILABILITY}
The data used in this paper are available from the the ESO and HST data archives.



\bibliographystyle{mnras}
\bibliography{main.bib} 








\bsp	
\label{lastpage}
\end{document}